\documentclass{elsart}
\usepackage{graphicx,amssymb,amsmath}
\usepackage{wrapfig,fancybox,bm,color}
\usepackage{extarrows}	

\begin{document}

\newtheorem{note2}[thm]{Note}
\newtheorem{method}{Method}
\journal{Int. J. Unconventional Computing}
\begin{frontmatter}
\title{A Dynamical Boolean Network}
\author{Genta Ito}
\ead{cxq02365@gmail.com}
\address{Maruo Lab., 500 El Camino Real \#302, Burlingame, CA 94010, United States.}

\begin{abstract}
We propose a Dynamical Boolean Network (DBN), which is a Virtual Boolean Network (VBN) whose set of states is fixed but whose transition matrix can change from one discrete time step to another. The transition matrix $T_{k}$ of our DBN for time step~$k$ is of the form $Q^{-1}TQ$, where $T$ is a transition matrix (of a VBN) defined at time step~$k$ in the course of the construction of our DBN and $Q$ is the matrix representation of some randomly chosen permutation $P$ of the states of our DBN. For each of several classes of such permutations, we carried out a number of simulations of a DBN with two nodes; each of our simulations consisted of 1,000 trials of 10,000 time steps each. In one of our simulations, only six of the 16 possible single-node transition rules for a VBN with two nodes were visited a total of 300,000 times (over all 1,000 trials). In that simulation, linearity appears to play a significant role in that three of those six single-node transition rules are transition rules of a Linear Virtual Boolean Network (LVBN); the other three are the negations of the first three.  We also discuss the notions of a Probabilistic Boolean Network and a Hidden Markov Model---in both cases, in the context of using an arbitrary (though not necessarily one-to-one) function to label the states of a VBN.
\end{abstract}
\end{frontmatter}

\section{Introduction}\label{SecIntro_p5}
One of the simplest and most common ways to categorize an entity as living or non-living is to stipulate that if we were to decompose that entity into its constituent parts and could then reconstruct the whole by assembling the parts, it is non-living; otherwise, it is living.  That is, a living thing is more than just the collection of its parts, but a non-living thing is not.  Then how is the gap between living and non-living things expressed?

R. Rosen~\cite{Rosen1,Rosen2,Rosen3}  claims that a central feature of living things is {\it complexity}, where a system is said to be {\it complex} if its behavior cannot be captured by models of that system; otherwise, that system is said to be {\it simple}. (That is, the complexity of a system depends primarily on the models of the system, and only secondarily on the system per se.)  In Rosen's description, use of the term {\it complex} in reference to a system is equivalent to use of the term {\it incomputable} or {\it notwell-formed} in reference to its models.

In his model of living things, Rosen considers a logical paradox (such as Russell's paradox~\cite{Lawvere1969}) as a {\it metabolism-repairsystem} ({\it MRsystem})~\cite{Rosen1}. An MR system consists of two sets (a set $X$ of {\it rawmaterials} and a set $Y$ of {\it behaviors}) and three functions (a {\it metabolicfunction}$f\in F=\mathrm{Hom}(X,$\,$Y)$, a {\it repairfunction}$g\in G=\mathrm{Hom}(Y,$\,$F)$, and a {\it replicationfunction}$h\in H=\mathrm{Hom}(F,$\,$G)$), where $g$ and $h$ are onto functions and $Y\simeq H$. (For sets $A,B,\ \mathrm{Hom}(A,B)$ denotes the set of all morphisms from $A$ to $B$~\cite{MacLane}.)

One remarkable feature of Rosen's model is the following:
\begin{enumerate}
\item[(i)]   There do not exist such onto functions $g\in G$ and $h\in H$.  If they are assumed to exist, we obtain a contradiction.
\end{enumerate}

\begin{figure}[h]
\begin{center}
\includegraphics[width=35mm,height=11.6mm]{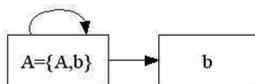}
\end{center}
\caption{Hyperset diagram of $A=\{A,\,b\}$}\label{FigHyperset1_p5}\end{figure}

One way to treat such a contradiction is to invoke hyperset theory~\cite{Kercel2003}.  A {\it hyperset} is defined as a graphable set---in particular, one that can be represented by a hyperset diagram, which is a digraph (directed graph) with the property that every node is either a set contained in the hyperset or an element of such a set, and every edge is directed from some set~$S$ contained in the hyperset to one of the elements of~$S$~\cite{Aczel1988,BarwiseEtchemendy1987}. Thus edge $a\rightarrow b$ expresses the relationship $a\ni b$, i.e., $b$ is an element of~$a$.  For example, the hyperset diagram of $A=\{A,\, b\}$ is shown in Fig.~\ref{FigHyperset1_p5}. There are two directed edges: $A\ni A$ and $A\ni b$.  A relationship of the form $A\ni A$ leads to the well-known contradiction known as Russell's paradox~\cite{Lawvere1969}; however, such a relationship is interpreted as just a loop structure in hyperset theory.  

\begin{figure}[h]
\begin{center}
\includegraphics[width=130mm,height=30.2mm]{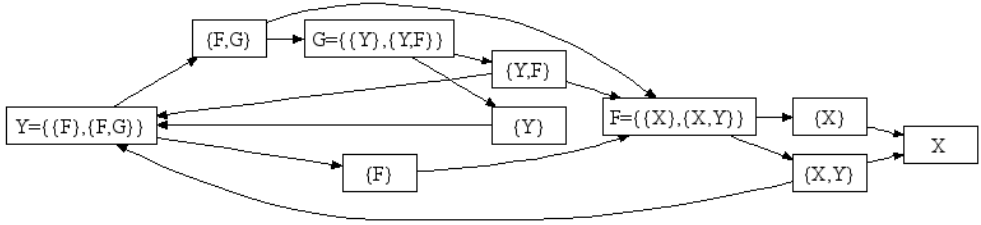}
\end{center}
\caption{The hypersetdiagram of MR}
\label{FigHyperset2_p5}\end{figure}

Property (i) can be expressed in hyperset theory by its way.  Let us represent $\mathrm{Hom}(A,\, B)$ by the set $\{\{A\},\,\{A,\, B\}\}$; thus $F,\, G$ and $H$ are represented by the sets $\{\{X\},\,\{X,\, Y\}\},\,\{\{Y\},\,\{Y,\, F\}\}$ and $\{\{F\},\,\{F,\, G\}\}$ respectively. Then we obtain the hyperset diagram of MR as shown in Fig.~\ref{FigHyperset2_p5}~\cite{ChemeroTurvey2006}. Accordingly, a system can be said to be complex if it cannot be well-formed in standard set theory but it can be well-formed in hyperset theory.

Matsuno and Gunji, in their theory of Internal Measurement~(IM)~\cite{Matsuno1,Matsuno2,GunjiItoKusunoki,GunjiHarunaSawa}, propose a different way, called a {\it perpetualequilibratingmechanism}.  They address property (i) as follows:
\begin{enumerate}
\item[(ii)]     Any solution to a problem (paradox) is inevitably a pseudo-solution, and the pseudo-solution at any given time step ``always'' triggers a problem that that is to be solved at the next time step. This leads to an evolutionary process that goes on forever and thereby allows for emergent properties.  Therefore, the paradox is treated as the force that drives the time evolution of the system.  
\end{enumerate}

In IM, the gap between the whole and the aggregation of the parts drives the time evolution of the dynamical system.  Therefore, IM can be understood as a proposal of a dynamical system which is based on the idea of a perpetual loop between two different logical layers, such as the parts and the whole, with a decomposition $\rho$ of the whole into the parts, and an integration $\rho^{-1}$ (as a pseudo-solution) of the parts into the whole.  Thus we may express an IM model in terms of a triad: the two different logical layers and a mediator/interface to adjust the two layers in an {\it inconsistent} manner~\cite{Matsuno2,GunjiHarunaSawa}.  IM claims that the relationship between the two layers (or the relationship between $\rho$ and $\rho^{-1}$) is not consistently determined, and for this reason they are perpetually changing relative to each another, where the term inconsistent means that we obtain a logical paradox if we assume consistency between them.

Any finite Boolean Network (BN) has a fixed transition matrix, hence a finite BN ultimately reaches an attractor and stays there (i.e., the computation ends). In section~\ref{SecVBN_p5}, we define a Virtual Boolean Network (VBN) as a BN in which the set of incoming nodes for every node is the entire set of nodes of the BN, and we show that every finite BN can be treated as a VBN. In section~\ref{SecExpansions_p5}, we introduce a labeling function to label the states of a VBN, and we discuss the Probabilistic Boolean Network (PBN) and the Hidden Markov Model (HMM) as extensions of a VBN. The purpose of an HMM is to find an optimal solution; in the case of a living thing, however, it is not always possible to find such a solution. Thus, in section~\ref{SecDBN_p5}, we abandon the attempt to find an optimal solution, and we introduce a Dynamical Boolean Network (DBN), which is a VBN whose set of nodes is fixed but whose transition matrix can change from one time step to another. Thus even a finite DBN does not necessarily have any attractors. In our DBN, a labeling function plays a role in maintaining a perpetual loop between the  of the VBN at different time steps.

In section~\ref{SecDBN_p5}, we present a construction of a DBN---a construction that incorporates stochasticity in a number of respects. In section \ref{SecSimulation_p5}, we present the results of a number of simulations of a DBN with two nodes. The purpose of each simulation was to investigate the set of rule vectors actually visited by the DBN, which is equivalent to the set of transition matrices actually visited by the DBN. In subsection~\ref{SubSecType1_p5}, we present the results of a simulation of a DBN based on the construction in section~\ref{SecDBN_p5}, which we call a type~1 simulation.  The simulations presented in the remaining three subsections of section~\ref{SecSimulation_p5}  use DBN constructions which are similar to the one in section~\ref{SecDBN_p5}, the only difference being that, for $k\ge 2$, we chose the transition matrix $T_{k}$ of the DBN for time step~$k$ to be $(Q_{k})^{-1}TQ_{k}$, where $T$ is the transition matrix in our original construction and $Q_{k}$ is the matrix representation of some permutation $P_{k}$ of the states of the DBN. Those three simulations correspond to three different classes of permutations. In our type~2 simulation, $P_{k}$ was randomly chosen from the set of permutations that consist of either the identity or a single 2-cycle or the product of two disjoint 2-cycles. In our type~3 simulation, $P_{k}$ was randomly chosen from the set of all 24 permutations of the four states of a DBN with two nodes. In our type~4 simulation, $P_{k}$ was constructed from a labeling function.

There are 81 rule vectors (out of a total of 256 possible rule vectors for a DBN with two nodes) that were not visited at all after the fifth time step of any trial in the type~1 and type~2 simulations.  However, all 256 rule vectors were visited in every trial of the type~3 simulation, as well as in slightly more than half of the trials of the type~4 simulation. Indeed, six of the 81 aforementioned rule vectors were the ones visited most frequently of all in the type~4 simulation. A total of six single-node transition rules are associated with those six rule vectors (each rule vector corresponds to an ordered pair of single-node rule vectors). Three of those six single-node transition rules are transition rules of a linear VBN (LVBN); the other three are the negations of the first three. Thus our conclusion is that the type~4 simulation presents new insight into linearity of a VBN.

\section{Boolean Networks and Virtual Boolean Networks}\label{SecVBN_p5}
A Boolean Network (BN)~\cite{Kauffman1,Kauffman2,Kauffman3}   is a digraph in which each vertex has an internal state and a transition rule.  The vertices of a BN are hereinafter referred to as nodes. Each internal state has a value of either 0 or 1 at any given time. The transition rule for a given node specifies the next state of that node as a function of the current internal states of its incoming (input) nodes.

A BN can be finite or infinite. The discussion of BN's in this paper is limited to finite BN's (i.e., those that have finitely many nodes). Some of the assertions made herein about BN's do not extend to the infinite case.

\begin{figure}[h]
\begin{center}
\includegraphics[width=100mm,height=29mm]{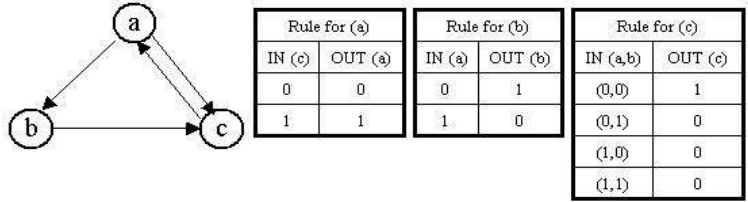}
\end{center}
\caption{Example of a BN}\label{Kauffman1_p5}
\end{figure}

An example of a BN is shown in Fig.~\ref{Kauffman1_p5}. There are three nodes ($a,b,c$), so the BN has a total of $8$ ($=2^{3}$) possible states:\[(0,0,0),(0,0,1),(0,1,0),(0,1,1),(1,0,0),(1,0,1),(1,1,0),(1,1,1)\]
 The set of incoming nodes for each node of a BN is specified by the digraph.  Here, the sets of incoming nodes for $a,\ b$, and $c$ are $\{c\},\,\{a\}$, and $\{a,b\}$, respectively.  Since nodes $a$ and $b$ have just one incoming node apiece, each of them has just two possible input states: $0$ and $1$.  Node $c$ has two incoming nodes, hence it has four possible input states: (0,0),(0,1),(1,0),(1,1).

Given the initial state of a BN, we obtain the time evolution of the system (the sequence of states visited) by repeated application of the transition rules.  Since the number of nodes is finite, the number of possible states is finite, so for every initial state the system evolves to some finite cyclic sequence of states, i.e., a finite sequence of states of the BN which is visited {\it ad infinitum from some time on, with no other states intervening}. Such a sequence is called an attractor.  Clearly, each state evolves to one and only one attractor. Thus the set of states of a BN can be partitioned in such a way that states in the same block of the partition evolve to the same attractor, and states in different blocks evolve to different attractors. 

\begin{figure}[h]
\begin{center}
\includegraphics[width=110mm,height=30mm]{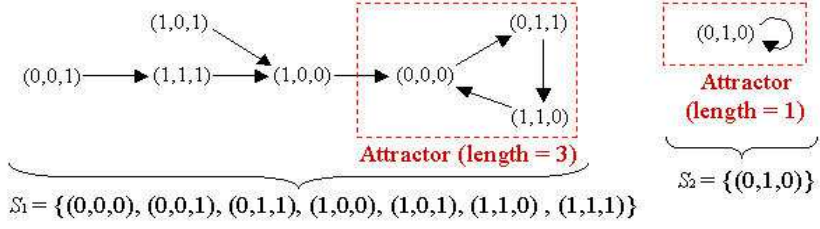}
\end{center}
\caption{Transition diagram of the BN in Fig.~\ref{Kauffman1_p5}}
\label{Kauffman2_p5}
\end{figure}

All possible states and attractors of the BN in Fig.~\ref{Kauffman1_p5}   are depicted in the transition diagram in Fig.~\ref{Kauffman2_p5}. In this particular BN, only four of the states are visited infinitely often, and there are two attractors: one of length 3 ($(0,0,0)\rightarrow(0,1,1)\rightarrow(1,1,0)$) and one of length 1 ($(0,1,0)$).  (The length of an attractor is the number of states in the cycle.) The set of states can be partitioned as $S_{1}\cup S_{2}$, where \[S_{1}=\{(0,0,0),(0,0,1),(0,1,1),(1,0,0),(1,0,1),(1,1,0),(1,1,1)\}\]
and\[S_{2}=\{(0,1,0)\},\]
 and every state in~$S_{1}$ evolves to the attractor of length 3.

A digraph with node set $V$ is specified by assigning to every node a set of incoming nodes.  Assigning a set of incoming nodes to a single node can be regarded as selecting an element of $\mathrm{Hom}(V,2)$, where $\mathrm{Hom}(A,B)$ denotes the set of all morphisms from $A$ to $B$~\cite{MacLane}. (Here, 2 denotes the set $\{0,1\}$.) The term {\it morphism} as used here is synonymous with {\it function}; hence $\mathrm{Hom}(V,2)$ is the set of all binary functions (functions whose values are elements of the set $2=\{0,1\}$) with $V$ as domain. Each such binary function can be represented as a binary sequence (a sequence of 0's and 1's) in which the $j$th term is the value of the function at node $j$. The set of incoming nodes for node $i$ is represented by the binary sequence in which the $j$th term is 1 if node $j$ is an incoming node for $i$, and 0 otherwise. Thus assigning to every node in~$V$ a set of incoming nodes (i.e., selecting an element of $\mathrm{Hom}(V,2)$ for every node) can be regarded as selecting an element of $\mathrm{Hom}(V,\mathrm{H}\mathrm{o}\mathrm{m}(V,2))$.

Alternatively, specification of a digraph can be regarded as selecting an element of $\mathrm{Hom}(V\times V,2)$, since 
\begin{equation}
\label{EqCCC_p5}
\mathrm{Hom}(V\times V,2)\simeq\mathrm{H}\mathrm{o}\mathrm{m}(V,\mathrm{H}\mathrm{o}\mathrm{m}(V,2))
\end{equation}
via the mapping of $\varphi\in\mathrm{Hom}(V\times V,2)$ to the element $\hat{\varphi}$ of $\mathrm{Hom}(V,\mathrm{H}\mathrm{o}\mathrm{m}(V,2))$ which is defined by $\hat{\varphi}(v_{1})=\varphi(v_{1},\mbox{---})$. For every $v_{1}\in V$, denote $\hat{\varphi}(v_{1})$ by $\hat{\varphi}_{v_{1}}$. Then $\hat{\varphi}_{v_{1}}\in \mathrm{Hom}(V,2)$ and, for every $v_{2}\in V$, 
\begin{equation*}
\hat{\varphi}_{v_{1}}(v_{2})=\varphi(v_{1},v_{2})
\end{equation*}
Note that 
\[\mbox{$\hat{\varphi}_{v_{1}}(v_2)=1\quad\longleftrightarrow\quad v_2$ is an input node for $v_1\quad\longleftrightarrow\quad\varphi(v_1,v_2)=1$}\]

The internal states of the nodes of a BN with node set~$V$ can be regarded as the components of a state~$\vec{x}=(x_{1},x_{2},\ldots,x_{|V|})$, where $|V|$ is the number of nodes in~$V$ and $x_{i}$ is the internal state of node $i$ (i.e., $x_{1},x_{2},\ldots,x_{|V|}\in 2$).

There are two ways to specify the transition rules for a BN. The first method is completely general.
\begin{method}\label{DefFirstRuleDescription_p5}      The transition rule $f_{i}$ for node $i$ assigns the next internal state (either 0 or 1) to node~$i$. The assignment is a function of the current internal states of the nodes in the set $W(i)$ of incoming nodes for $i$.
  
Let $W(i)=\{i_{1},i_{2},\ldots,i_{|W(i)|}\}$, where $|W(i)|$ is the number of nodes in~$W(i)$. Then\[f_{i}\in\mathrm{Hom}(\mathrm{Hom}(W(i),2),2),\]
where \[\mathrm{Hom}(W(i),2)=\{(x_{i_{1}},x_{i_{2}},\ldots,x_{i_{|W(i)|}}):x_{i_{1}},x_{i_{2}},\ldots,x_{i_{|W(i)|}}\in 2\}\]

The transition rules for the nodes of a BN can be regarded as the components of a rule vector~$\vec{f}$:\[\vec{f}=(f_{1},f_{2},\ldots,f_{|V|})\in\prod_{j\in V}\mathrm{Hom}(\mathrm{Hom}(W(j),2),2)\]
\end{method}

The second method of specifying the transition rules applies only to a BN for which $W(i)=V$ (hence $\overline{W(i)}=\emptyset$) for every node $i$, where $\overline{W(i)}$ denotes $V-W(i)$. We will refer to such a BN as a Virtual Boolean Network (VBN).

\begin{method}\label{DefSecondRuleDescription_p5}    If $W(i)=V$ for every $i$, then the transition rules $f_{i}$ for the individual nodes can be incorporated into a single transition rule $F$:
\begin{equation}
\label{RuleSpecification3}
\mbox{$F\in \mathrm{Hom}(V, \mathrm{Hom}(\mathrm{Hom}(V,2),2)),\ \ F(i)\equiv f_i\in \mathrm{Hom}(\mathrm{Hom}(V,2),2)$}
\end{equation}
\end{method}

For node $i$, denote $F(i)$ by $F_{i}$. For $j\in V$, we will say that node $j$ is {\it virtually disconnected} from node $i$ if $F_{i}(\ldots,y_{j},\ldots)$ is invariant with respect to substitution of input $y_{j}$ with $\sim y_{j}$, where\[y_{j}=0\implies\sim y_{j}=1,\qquad y_{j}=1\implies \sim y_{j}=0\]
We will say that the nodes which are not virtually disconnected from node $i$ are {\it virtual incoming nodes} for node~$i$, and we will denote the set of virtual incoming nodes by $VW(i)$; hence the set of nodes which are virtually disconnected from node $i$ is $\overline{VW(i)}=V-VW(i)$.  

Any BN can be regarded as a VBN, by defining $F_{i}$ in such a way that it extends $f_{i}$ and all the nodes in $\overline{W(i)}$ are virtually disconnected from $i$.

\begin{exmp}\label{Ex1_p5}
Let $V=\{a,b,c\}$, and fix a node $i$. Suppose that $W(i)=\{a\}$ and $\overline{W(i)}=\{b,c\}$, and that the transition rule $f_{i}$ for node $i$ is
\begin{equation*}
f_{i}=\left[\begin{array}{lc|cc}
\emph{In}&(x_{a})&0&1\\\hline
\emph{Out}&(x_{i})&1&0\end{array}\right]
\end{equation*}

The domain of rule $f_{i}$ is just the set of states of node $a$, but $f_{i}$ can be embedded into rule $F_{i}$ defined by 
\begin{equation*}
F_{i}=\left[\begin{array}{l}\begin{array}{lc|cccc}
\emph{In}&(x_{a},x_{b},x_{c})&(0,0,0)&(0,0,1)&(0,1,0)&(0,1,1)\\\hline
\emph{Out}&(x_{i})&1&1&1&1\end{array}\\[.4in]\begin{array}{lc|cccc}
\emph{In}&(x_{a},x_{b},x_{c})&(1,0,0)&(1,0,1)&(1,1,0)&(1,1,1)\\\hline
\emph{Out}&(x_{i})&0&0&0&0\end{array}
\end{array}\right]
\end{equation*}

The domain of $F_{i}$ is the set of states of all the nodes. Note that, for all $x_{a},x_{b},x_{c}\in\{0,1\}$, the embedding satisfies
\begin{equation*}
F_{i}(x_{a},x_{b},x_{c})=F_{i}(x_{a},\sim x_{b},x_{c})=F_{i}(x_{a},x_{b},\sim x_{c})=F_{i}(x_{a},\sim x_{b},\sim x_{c})
\end{equation*}
Hence nodes $b$ and $c$ are virtually disconnected from node~$i$.
\end{exmp}

We can translate a specification of a VBN into a specification of a BN, though the resulting BN is not necessarily unique (since $W(i)$ can be any superset of $VW(i)$). For node~$i$ in Example~\ref{Ex1_p5}, we could define $W(i)$ as $\{a,b\}$, and the transition rule $g_{i}$ for node $i$ as

\begin{equation*}
 g_{i}=\left[\begin{array}{lc|cccc}\mbox{In}&(x_{a},x_{b})&(0,0)&(0,1)&(1,0)&(1,1)\\\hline
\mbox{Out}&(x_{i})&1&1&0&0\\\end{array}\right]
\end{equation*}

Note that, for all $x_{a},x_{b}\in\{0,1\}$,\[f_{i}(x_{a})=g_{i}(x_{a},x_{b})=g(x_{a},\sim x_{b})\]
Furthermore, both $f_{i}$ and $g_{i}$ embed into the transition rule $F_{i}$ given in Example~\ref{Ex1_p5}.

The transition diagram of a BN is a digraph in which the vertices are the states of the BN and every vertex has just one out-going edge. If the set of nodes of the BN is~$V$, then the BN has a total of $2^{|V|}$ states. Thus the transition diagram can be represented as a $2^{|V|}\times 2^{|V|}$ {\it adjacency matrix} . The adjacency matrix $T$  of any digraph is a binary matrix (i.e., a matrix of 0's and 1's) such that $T_{ij}=1$ if there is an out-going edge from the vertex labeled~$i$ to the vertex labeled~$j$ (and $T_{ij}=0$ otherwise). Since each state of a BN has just one out-going edge, exactly one of the entries in each row of the adjacency matrix of a BN is a 1. In the case of a BN, the adjacency matrix is also called the {\it transition matrix}.

\begin{prop}\label{PropRules2TRANS1_p5}
The transition matrix $T$ of a BN with vertex set~$V$ can be obtained from the rule vector $\vec{f}$; however, $\vec{f}$ cannot be recovered from $T$.  
\end{prop}
{\bf Proof}\,\,\,      ($\vec{f}\Rightarrow T$): Label the rows of~$T$ with the states of the BN, and use the same labels (in the same order) for the columns of~$T$. Let $\vec{x}=(x_{1},x_{2},\ldots,x_{|V|})$ be a state of the BN. Then, in the row of~$T$ which is labeled with state~$\vec{x}$, the unique 1 is located in the column labeled with state~$\vec{y}=(y_{1},y_{2},\ldots,y_{|V|})$, where, for every~$i,\ \ y_{i}$~is computed by applying transition rule $f_{i}$ to~$\vec{x}$.  \\
($T\not\Rightarrow\vec{f}$): To determine transition rule~$f_{i}$, its domain (namely, $W(i)$) must be known; however, $W(i)$ cannot be derived from~$T$.  $\blacksquare$

The following proposition follows from the fact that, for every node in a VBN with node set~$V$, the set of incoming nodes is $V$.

\begin{prop}\label{PropRules2TRANS2_p5}
In a VBN, the transition matrix $T$ can be derived from the transition rule~$F$, and $F$~can be derived from~$T$.
\end{prop}
{\bf Proof}\,\,\,      
By Proposition~\ref{PropRules2TRANS1_p5}, it suffices to prove that $T\Rightarrow F$. Assume that the rows and columns of~$T$ are labeled with the states of the VBN, as indicated in the proof of Proposition~\ref{PropRules2TRANS1_p5}, and let $i\in\{1,2,\ldots,|V|\}$. The output of transition rule~$F_{i}$, which is the next internal state of node~$i$, is a function of the current state of the BN, so let $\vec{x}$ be a state of the VBN. Furthermore, let $j$~be such that, in the row of~$T$ which is labeled with state~$\vec{x}$, the unique 1 is located in column~$j$, and let $\vec{y}=(y_{1},y_{2},\ldots,y_{|V|})$ be the state that labels column~$j$. Then $y_{i}$ is the output of $F_{i}$ that corresponds to input~$\vec{x}.\ \blacksquare$

From this point on, we assume that every BN is a VBN, so we need not specify the set of incoming nodes for any node.  Every state $\vec{x}$ of a VBN with node set~$V$ is an element of $\mathrm{Hom}(V,2)$, and the transition rule $F$ for such a VBN is an element of  
$\mathrm{Hom}(V,\mathrm{Hom}(\mathrm{Hom}(V,2),2))$.  

\begin{defn}\label{DefLBN_p5}
A VBN is said to be a \emph{linear VBN (LVBN)} if, for every node~$i$, transition rule $F_{i}$ is of the form
\[F_{i}(\vec{x})=(\vec{x}\cdot\vec{v}_{i})\bmod 2,\]
where $\vec{x}$ is the current state of the VBN and $\vec{v}_{i}$ is a vector whose components are 0's and 1's. (Clearly, the number of components of~$\vec{v}_{i}$ must be equal to the number of nodes in the VBN.) 
\end{defn}

\begin{prop}\label{PropMBN1_p5}
In an LVBN, node~$i$ is virtually disconnected from node~$j$ if and only if the $j$th component of the vector $\vec{v}_{i}$ in Definition~\ref{DefLBN_p5}   is 0.
\end{prop}
{\bf Proof}\,\,\,The output of transition rule~$F_{i}$ (namely, $(\vec{x}\cdot\vec{v}_{i})\bmod 2$) is invariant under the transformation that consists of replacing the current internal state~$x_{j}$ of node~$j$ with~$\sim x_{j}$ (and leaving the current internal states of the other nodes unchanged) if and only if the $j$th component of $\vec{v}_{i}$ is 0.  $\blacksquare$

For a VBN with set of states~$S$ and node set~$V$, the number of possible single-node transition rules is $2^{|S|}=2^{2^{|V|}}$. In the case of a VBN with two nodes, this is $2^{2^{2}}=16$.

The output of each of the 16 possible single-node transition rules for a VBN with two nodes is given in Table~\ref{TableRule_p5} as a function of the input state of the VBN; tabulated along with the output of each transition rule is the number $n$ of virtual incoming nodes for each of the node(s) to which that rule would be applied.  There are only four possible transition rules for a node of an LVBN with two nodes (namely, the rules which are presented in boldface in the table). Those four transition rules (1, 7, 11, and 13) correspond to the vectors $\vec{v}=(0,0),\ \vec{v}=(1,1),\ \vec{v}=(0,1)$, and $\vec{v}=(1,0)$, respectively, where $\vec{v}$ is as in Definition~\ref{DefLBN_p5}. If nodes 1 and 2 of an LVBN are $a$ and $b$ (in that order), then the sets of virtual incoming nodes  for each of the nodes to which rules 1, 7, 11, and 13 would be applied are $\emptyset,\ \{a,b\},\ \{b\}$, and $\{a\}$, respectively.

\begin{table}[htbp]
\caption{Outputs of the transition rules for a VBN with two nodes.}\label{TableRule_p5}
\scriptsize
\begin{tabular}{|c|cccccccccccccccc|c|}\hline \nonumber
Input$\backslash$Rule no. 
&{\bf 1}&  2 & 3 & 4 &5 & 6 & {\bf 7} & 8 & 9 &10&{\bf 11}&12&{\bf 13} &14&15&16 \\ \hline
(0,0) & {\bf 0} & 1 & 0 & 1 & 0 & 1 & {\bf 0} & 1 & 0 & 1 & {\bf 0} & 1 & {\bf 0} & 1 & 0 & 1 \\ \hline
(0,1) & {\bf 0} & 0 & 1 & 1 & 0 & 0 & {\bf 1} & 1 & 0 & 0 & {\bf 1} & 1 & {\bf 0} & 0 & 1 & 1 \\ \hline
(1,0) & {\bf 0} & 0 & 0 & 0 & 1 & 1 & {\bf 1} & 1 & 0 & 0 & {\bf 0} & 0 & {\bf 1} & 1 & 1 & 1 \\ \hline
(1,1) & {\bf 0} & 0 & 0 & 0 & 0 & 0 & {\bf 0} & 0 & 1 & 1 & {\bf 1} & 1 & {\bf 1} & 1 & 1 & 1 \\ \hline
$n$    & {\bf 0} & 2 & 2 & 1 & 2 & 1 & {\bf 2} & 2 & 2 & 2 & {\bf 1} & 2 & {\bf 1} & 2 & 2 & 0 \\ \hline
\end{tabular}
\end{table}

\section{Labeling Functions}\label{SecExpansions_p5}
In section~\ref{SecVBN_p5}, we found that the transition matrix~$T$ of a BN has the property that exactly one entry in each row is a 1.  We will use the term {\it Boolean matrix} for every square binary matrix that has exactly one ``1'' in each row.  By Proposition~\ref{PropRules2TRANS2_p5}, for every $n\ge 1$ there is a natural one-to-one correspondence between the set of all $2^{n}\times 2^{n}$ Boolean matrices and the set of all VBN's with node set $\{1,2,\ldots,n\}$, provided that the rows and columns of all the Boolean matrices are labeled in the same way (e.g., with the states of such a VBN) and in the same order. 

For a VBN with node set~$V$, let $S$ be the set of states, let\[\Lambda=\{1,2,3\ldots,2^{|V|}\},\]
 and let $\Xi:S\rightarrow\Lambda$ be a function that assigns a label to each state.

For any state $s_{0}$ of the VBN, let\[ s_{0},\, s_{1},\, s_{2},\,\ldots\]
be a sequence of states such that $T(s_{i})=s_{i+1}$ for every $i$, where $T$ is the transition matrix of the VBN and $T(s_{i})$ is the unique output state for input state~$s_{i}$. Furthermore, let
\[\alpha_{0},\,\alpha_{1},\,\alpha_{2},\,\ldots,\]
be the sequence of labels of the corresponding output states. Since $T$ is a transition matrix of a VBN, there is just one such sequence of states with $s_{0}$ as first term (hence just one such sequence of labels with $\alpha_{0}$ as first term).

\begin{figure}[h]
\begin{center}
\includegraphics[width=20mm,height=12.8mm]{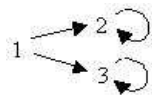}
\end{center}
\caption{Digraph that has two sequences of labels with first term 1}
\label{FigCounterExample_p5}\end{figure}

There are digraphs with underlying set~$S$ whose adjacency matrix is not Boolean. Thus for some state~$s_{0}$, there are at least two different sequences of states with first term~$s_{0}$ (hence there are at least two different sequences of labels with first term~$\alpha_{0}$).  For the digraph shown in Fig.~\ref{FigCounterExample_p5}, for example, there are two sequences of labels with first term 1: 1,2,2,2,$\ldots$ and 1,3,3,3,$\ldots$. The adjacency matrix of that digraph is not Boolean, because state 1 can make a transition to either state 2 or state 3, so the digraph is not a transition diagram of any BN.

\begin{figure}[h]
\begin{center}
\includegraphics[width=30mm,height=8.6mm]{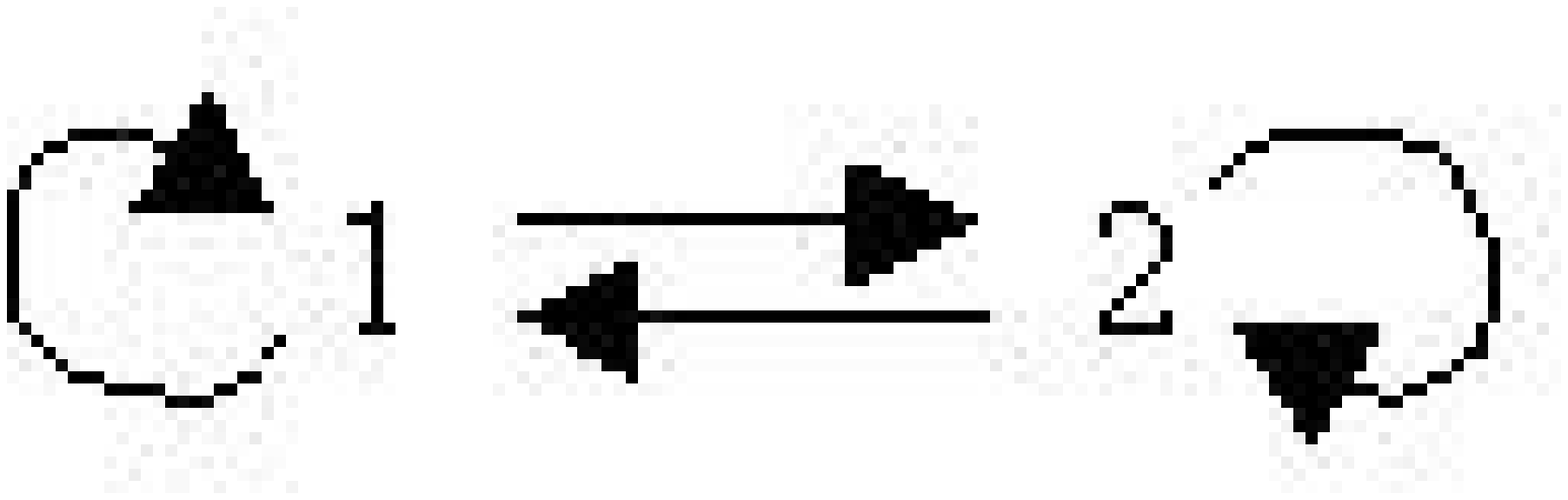}
\end{center}
\caption{Output digraph for the sequence of labels in (\ref{Eq12122111_p5})}
\label{Fig12122111_p5}\end{figure}

For any given sequence of labels, there exists a unique digraph, which we will call the {\it output digraph}. For example, the output digraph for the finite sequence of labels
\begin{equation}
\label{Eq12122111_p5}
1,2,1,2,2,1,1,1
\end{equation}
 is shown in Fig.~\ref{Fig12122111_p5}. The adjacency matrix of that digraph is not Boolean, since state 1 can make a transition to either state 1 or state 2 (as can state 2). Therefore, the digraph is not a transition diagram of any BN.

There are at least four ways to treat a finite sequence of labels of vertices of a digraph, such as the sequence in (\ref{Eq12122111_p5}). In each of those four methods, the labels are treated as the values of a deterministic or stochastic labeling function~$\Xi$ whose domain is a subset of the set of states of some VBN or Probabilistic Boolean Network (PBN). In a PBN, the transition rules are regarded as random variables, and the transition matrix is a probabilistic transition matrix~\cite{Shmulevich1,Shmulevich2}. 

\begin{figure}[h]
\begin{center}
\includegraphics[width=120mm,height=7.4mm]{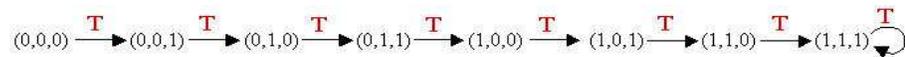}
\end{center}
\caption{Transition diagram of a VBN with a sequence of distinct states that can be associated with the sequence of labels in (\ref{Eq12122111_p5})}
\label{FigLineVBN_p5}\end{figure}

The first way is to use a VBN and a deterministic labeling function. For example, we could use a VBN in which there is a sequence $s_{0},s_{1},s_{2},\ldots,s_{n-1}$ of distinct states such that $T(s_{i})=s_{i+1}$ for every $i<n-1$, where $n$ is the number of terms in the given sequence of labels and $T$ is the transition matrix of the VBN. The sequence in~(\ref{Eq12122111_p5}) has 8 labels, so a VBN with a sequence of 8 distinct states will suffice (e.g., the VBN with 3 nodes whose transition diagram is shown in Fig.~\ref{FigLineVBN_p5}).

Once the VBN with the sequence of $n$ distinct states is constructed, we label state $s_{i}$ with the original label $\alpha_{i}$, i.e., we define the following many-to-one deterministic labeling function $\Xi$:
\begin{equation*}
\Xi=\left[\begin{array}{cccccccc}
(0,0,0)&(0,0,1)&(0,1,0)&(0,1,1)&(1,0,0)&(1,0,1)&(1,1,0)&(1,1,1)\\\hline
1&2&1&2&2&1&1&1\end{array}
\right]
\end{equation*}

Though we can recover the output digraph from the original sequence of labels, the fact that a sequence of distinct states of a VBN was used as the domain of the labeling function~$\Xi$ is lost, so this way of treating a finite sequence of labels of vertices of a digraph is rather artificial.  

\begin{figure}[h]
\begin{center}
\includegraphics[width=51mm,height=20mm]{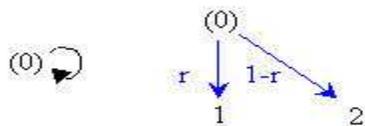}
\end{center}
\caption{Representation of the digraph in Fig.~\ref{Fig12122111_p5} by the simplest VBN (left) and a one-to-many (stochastic) labeling function $E$ (right)}
\label{FigOnetoManyE_p5}\end{figure}

The second way is to use a VBN and a stochastic labeling function $\Xi$.  For the sequence in~(\ref{Eq12122111_p5}), we could use a VBN with just a single state, (0), as shown in the left half of Fig.~\ref{FigOnetoManyE_p5},  and the stochastic labeling function $\Xi$ portrayed in the right half of the figure, which assigns the label ``1'' with some non-zero probability $r$ and the label ``2'' with probability $1-r$.

We can recover any output digraph in this way, that is, by choosing a suitable stochastic labeling function $\Xi$ on the set of states of an arbitrary VBN.  However, just as with the first method, the fact that there is an underlying VBN is lost.

The third way to treat a finite sequence of labels is to use a PBN and a deterministic labeling function $\Xi$.  For the sequence of labels in (\ref{Eq12122111_p5}), for example, we construct the digraph in Fig.~\ref{Fig12122111_p5}    and then define a stochastic transition matrix~$T$ by assigning a weight (probability) to each edge. For example, we could use the matrix\[T=\left(\begin{array}{cc}
0.2&0.8\\
0.4&0.6\end{array}\right)\]

\begin{figure}[h]
\begin{center}
\includegraphics[width=39mm,height=10mm]{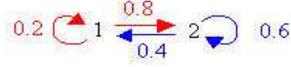}
\end{center}
\caption{PBN that can be associated with the sequence of labels in~(\ref{Eq12122111_p5})}
\label{Fig12122111PBN_p5}\end{figure}

If we use the deterministic function~$\Xi$ that labels the first state of the PBN with ``1'' and the second state with ``2'', we obtain the PBN in Fig.~\ref{Fig12122111PBN_p5}.

The fourth way is to use a PBN and a stochastic labeling function. Thus we define stochastic matrices~$T$ and $\Xi$. If $S$ is the set of states of the PBN and $\Lambda$ is the set of labels, then the following hold for every state~$s$: 

\[\sum_{\alpha\in\Lambda}\Xi_{s\alpha}=1=\sum_{s^{\prime}\in S}T_{ss^{\prime}},\]
where $T_{ss^{\prime}}$ is the probability that the PBN makes a transition from state $s$ to state~$s^{\prime}$ and $\Xi_{s\alpha}$ is the probability that state $s$ is labeled with~$\alpha$.

Such an entity is an example of what is known as a Hidden Markov Model (HMM), a statistical model in which a system is assumed to be characterized by a stochastic transition matrix $T$ whose rows and columns are labeled with the states of the system and a stochastic matrix $\Xi$ (known as the {\it emission matrix}) whose rows and columns are labeled with the values of some variable $\alpha$ that takes only finitely many different values. Typically, the values of~$\alpha$ are observable, but the matrices~$T$ and~$\Xi$ are unknown and the states of the system are hidden (not directly observable).  

The following is an iterative process that, given a positive integer~$n$ and a sequence $\vec{\alpha}^{(n)}=(\alpha_{0},\ldots,\alpha_{n-1})$ of empirically determined values of~$\alpha$, approximates the matrices~$T$ and~$\Xi$ and finds an appropriate sequence $\vec{s}^{(n)}=(s_{0},\ldots,s_{n-1})$ of states of the system such that state $s_{i}$ can be assigned the value~$\alpha_{i}$.

\begin{enumerate}
\item[\textit{Step~1}]      Define a nonnegative function~$||\cdot||$ such that, for all pairs $(\vec{\alpha},\vec{\beta})$ of finite sequences of values of~$\alpha$ with equal numbers of terms, $||\vec{\beta}-\vec{\alpha}||$ is a measure of the difference between $\vec{\alpha}$ and $\vec{\beta}$. Then set an upper limit~$\epsilon$ ($>0$) on the acceptable value of $||\cdot||$ for such pairs of sequences.
\item[\textit{Step~2}]      Input a positive integer~$n$ and an empirically determined sequence $\vec{\alpha}^{(n)}=(\alpha_{0},\ldots,\alpha_{n-1})$ of values of~$\alpha$, and randomly choose a sequence $\vec{s}^{(n)}=(s_{0},\ldots,s_{n-1})$ of states of the system.
\item[\textit{Step~3}]      Input the sequences $\vec{\alpha}^{(n)}$ and $\vec{s}^{(n)}$ to an algorithm~$\rho$ that outputs the most likely pair $(\Xi,T)$ of stochastic matrices for the system:\[\rho(\vec{\alpha}^{(n)},\vec{s}^{(n)})\rightarrow(\Xi,\, T)\]
\item[\textit{Step~4}]      Input the sequence $\vec{\alpha}^{(n)}$ and the estimated pair $(\Xi,\, T)$ to an algorithm~$\sigma$ that outputs the most likely sequence $\vec{t}^{(n)}=(t_{0},\ldots,t_{n-1})$ of states of the system:\[\sigma(\vec{\alpha}^{(n)},\Xi,T)\rightarrow\vec{t}^{(n)}\]
\item[\textit{Step~5}]      Input the sequence $\vec{t}^{(n)}$ and the estimated pair $(\Xi,\, T)$ to an algorithm~$\gamma$ that outputs the most likely sequence $\vec{\beta}^{(n)}=(\beta_{0},\ldots,\beta_{n-1})$ of values of~$\alpha$:\[\gamma(\vec{t}^{(n)},\Xi,T)\rightarrow\vec{\beta}^{(n)}\]
\item[\textit{Step~6}]      Compute the quantity\[\delta\equiv||\vec{\beta}^{(n)}-\vec{\alpha}^{(n)}||\]
If $\delta<\epsilon$, output $\vec{s}^{(n)}$ as the sequence of states that corresponds to the sequence $\vec{\alpha}^{(n)}$ of values of~$\alpha$, and then halt. Otherwise, set $\vec{s}^{(n)}$ to $\vec{t}^{(n)}$ and go to Step~3. 
\end{enumerate}

The purpose of HMM is to find an optimal solution with $\delta<\epsilon$. In a living thing, it is not always possible to find such a solution. Thus at this point we will terminate our discussion of the four methods of treating a finite sequence of labels of a digraph which we have presented in this section.  In section~\ref{SecDBN_p5}, we will introduce a fifth method (namely, a Dynamical Boolean Network (DBN)) as an alternative to all four of them, which does not depend on satisfying the condition $\delta<\epsilon$.

\section{Dynamical Boolean Networks}\label{SecDBN_p5}
In this section we construct a Dynamical Boolean Network (DBN), which we define as a VBN whose transition matrix can change from one discrete time step to another. The number of nodes of the DBN remains constant throughout; hence the set of states is invariant.

Our method of constructing a DBN is completely general. We will apply it to an example of a DBN with 2 nodes, so there are 4 states.  At each time step, we will consider a sequence of 5 states, hence at least one of the 4 states of the DBN is repeated in the sequence.

{\bf [Step 1]}  Let $k$ index the time steps, and initialize~$k$ to~$1$. Fix the number~$\mu$ of nodes of the DBN, and choose the initial labeling function~$\Xi_{1}:S\rightarrow\Lambda$ at random (where $S$ is the set of states of the DBN and $\Lambda=\{1,2,3,\ldots,2^{\mu}\}$). Next, randomly choose the initial transition matrix $T_{1}$, and a sequence of states $\vec{S}_{1}$ of the VBN whose transition matrix is~$T_{1}$. Then determine the sequence of labels $\vec{\alpha}_{1}$ which corresponds to $\vec{S}_{1}$ via the function~$\Xi_{1}$, and construct the output digraph $g=(V,E)$ for~$\vec{\alpha}_{1}$, where $V$ and $E$ are the vertex set and edge set, respectively.

\begin{figure}[h]
\begin{center}
\includegraphics[width=100mm,height=19.4mm]{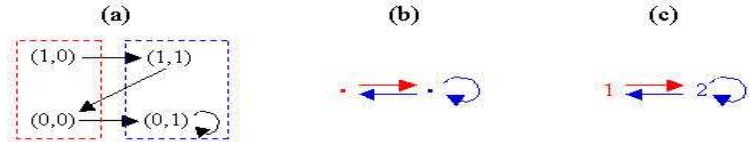}
\end{center}
\caption{(a) Initial transition diagram of a DBN which corresponds to transition matrix~$T_{1}$ and yields the sequence of states $\vec{S}_{1}$. (b) Output digraph for the sequence of labels~$\vec{\alpha}_1$, without labels.  (c) Output digraph for~$\vec{\alpha}_1$, with labels.}
\label{FigDBN1_p5}\end{figure}

In our example, $\mu=2$. For our initial labeling function, we will choose the many-to-one function
\begin{equation*}
\Xi_{1}=\left[\begin{array}{cccc}(0,0)&(0,1)&(1,0)&(1,1)\\\hline 1&2&1&2\end{array}\right]
\end{equation*}
Such a labeling function partitions the set of states into groups and assigns a different label to each group. For our initial transition matrix, we will choose
\[T_{1}=\left(\begin{array}{cccc}0&1&0&0\\0&1&0&0\\0&0&0&1\\1&0&0&0\\\end{array}\right),\]
which corresponds to the transition diagram in Fig.~\ref{FigDBN1_p5}(a), and we will use \[\vec{S}_{1}=(1,0),(1,1),(0,0),(0,1),(0,1)\]
as our sequence of states.

The sequence of labels $\vec{\alpha}_{1}$ which corresponds to $\vec{S}_{1}$ via the function~$\Xi_{1}$ is\[\vec{\alpha}_{1}=1,2,1,2,2\]
The unlabeled and labeled versions of the output digraph for~$\vec{\alpha}_{1}$ are shown in Figs.~\ref{FigDBN1_p5}(b) and~\ref{FigDBN1_p5}(c), respectively.  Clearly, the adjacency matrix of the output digraph is not Boolean.

{\bf [Step 2]}  Use~$g$ to construct a set $\mathcal{G}$ of digraphs in which every vertex has just one out-going edge (hence $\mathcal{G}$ is a set of digraphs whose adjacency matrices are Boolean). If every vertex of~$g$ has just one out-going edge, let $\mathcal{G}=\{g\}$ and go to Step~3. Otherwise, proceed as follows:  

\begin{itemize}
\item[(i)]      Select a vertex $x_{1}$ that has out-degree at least 2, let $d$ be the out-degree of $x_{1}$, let $y_{1},y_{2},\ldots,y_{d}$ be the vertices to which $x_{1}$ is connected via an out-going edge, and let  $x_{2},x_{3},\ldots,x_{d}$ be new vertices (not in $V$). If there is a loop at $x_{1}$ (i.e., if $(x_{1},x_{1})\in E$), then, for the sake of convenience, set $y_{1}$ to $x_{1}$. Assign entirely new labels (labels that have not yet appeared in the construction of our DBN) to the $d-1$ new vertices.

In the output digraph in Fig.~\ref{FigDBN1_p5}(c), there is just one vertex with at least two out-going edges (namely, the vertex labeled with 2), so $x_{1}$ is that vertex.  There are two out-going edges from $x_{1}$: one to $x_{1}$ itself (which is therefore also $y_{1}$) and one to the vertex labeled with 1 (which is thus $y_{2}$). Hence $d=2$, so we add $d-1=1$ new vertex, $x_{2}$, which we will label with $2^{\prime}$.

\item[(ii)]      For every one-to-one function \[\phi:\{x_{1},x_{2},\ldots,x_{d}\}\rightarrow\{y_{1},y_{2},\ldots,y_{d}\},\]
 construct the new digraph $g_{\phi}=(V_{\phi},E_{\phi})$ that has vertex set $V_{\phi}=V\cup\{x_{2},\ldots,x_{d}\}$ and edge set 
\begin{eqnarray*}
E_{\phi}=&\bigl(E-&\{(x_{1},y_{1}),(x_{1},y_{2}),\ldots,(x_{1},y_{d})\}\bigr)\\&&\bigcup\,\{(x_{1},\phi(x_{1})),(x_{2},\phi(x_{2})),\ldots,(x_{d},\phi(x_{d}))\}
\end{eqnarray*}
The number of digraphs~$g_{\phi}$ constructed here is $d!$.

In our example, we construct $d!=2!=2$ new digraphs, $g_{1}$ and $g_{2}$, one for each one-to-one function $\phi:\{x_{1},x_{2}\}\rightarrow\{y_{1},y_{2}\}$, as shown in Fig.~\ref{FigDBN1a_p5}.

\begin{figure}[h]
\begin{center}
\includegraphics[width=50mm,height=19.2mm]{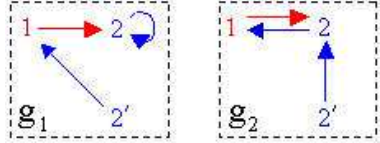}
\end{center}
\caption{The two digraphs constructed in Step~2(ii) in our example}
\label{FigDBN1a_p5}\end{figure}

\item[(iii)]      
If $g$ has a loop at $x_{1}$, construct additional new digraphs: For every $i\in\{2,\ldots,d\}$ and every one-to-one function \[\phi_{i}:\{x_{1},x_{2},\ldots,x_{d}\}\rightarrow\{x_{i},y_{2},\ldots,y_{d}\},\]
construct the new digraph $g_{\phi_{i}}=(V_{\phi_{i}},E_{\phi_{i}})$ that has vertex set $V_{\phi_{i}}=V\cup\{x_{2},\ldots,x_{d}\}$ and edge set
\begin{eqnarray*}
E_{\phi_{i}}=&\bigl(E-&\{(x_{1},y_{1}),(x_{1},y_{2}),\ldots,(x_{1},y_{d})\}\bigr)\\&&\bigcup\,\{(x_{1},\phi_{i}(x_{1})),(x_{2},\phi_{i}(x_{2})),\ldots,(x_{d},\phi_{i}(x_{d}))\}
\end{eqnarray*}
For every $i\in\{2,\ldots,d\}$, the number of digraphs $g_{\phi_{i}}$ is $d!$, hence the sum of the numbers of digraphs $g_{\phi_{i}}$ for all $i\in\{2,3,\ldots,d\}$ is $(d-1)\cdot d!$. Note that every new digraph constructed in Step~2(ii) or Step~2(iii) has the property that if we identify all the vertices $x_{1},x_{2},\ldots,x_{d}$, then that digraph collapses to $g$.

Since the output digraph $g$ in our example has a loop at $x_{1}$, we construct $(d-1)\cdot d!=(2-1)\cdot 2!=2$ additional new digraphs, $g_{1}^{\prime}$ and $g_{2}^{\prime}$, one for each one-to-one function $\phi_{2}:\{x_{1},x_{2}\}\rightarrow\{x_{2},y_{2}\}$, as shown in Fig.~\ref{FigDBN1b_p5}.

\begin{figure}[h]
\begin{center}
\includegraphics[width=50mm,height=19.2mm]{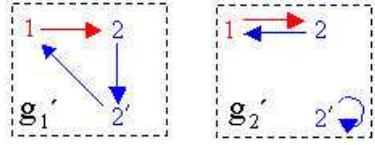}
\end{center}
\caption{The two digraphs constructed in Step~2(iii) in our example}
\label{FigDBN1b_p5}\end{figure}

\item[(iv)]      If each of the new digraphs just constructed in Step~2(ii) or Step~2(iii) has only one out-going edge, let $\mathcal{G}$ be the set that consists of all of those digraphs and go to Step~3. Otherwise,  go to Step~2(i) and apply this procedure in parallel to all of the new digraphs just constructed (by treating each of them as~$g$).

Since $x_{1}$ is the only vertex in our output digraph $g$ with at least two out-going edges, every vertex of each of the four new digraphs constructed in Step~2(ii) or Step~2(iii) has just one out-going edge, so we go to Step~3.
\end{itemize}

{\bf [Step~3]} The digraphs in~$\mathcal{G}$ all have the same number~$\lambda$ of vertices, and their adjacency matrices are all Boolean. If $\lambda<2^{\mu}$, let $\nu=2^{\mu}-\lambda$ and, for every digraph~$h\in\mathcal{G}$, construct all digraphs that are formed from~$h$ by adding $\nu$ new vertices $z_{1},z_{2},\ldots,z_{\nu}$ and assigning just one out-going edge to each new vertex. Then all of the resulting digraphs have $2^{\mu}$ vertices, and their adjacency matrices are transition matrices of actual VBN's. We will call each of those digraphs a {\it pseudo-transition diagram}.

\begin{figure}[h]
\begin{center}
\includegraphics[width=120mm,height=79.6mm]{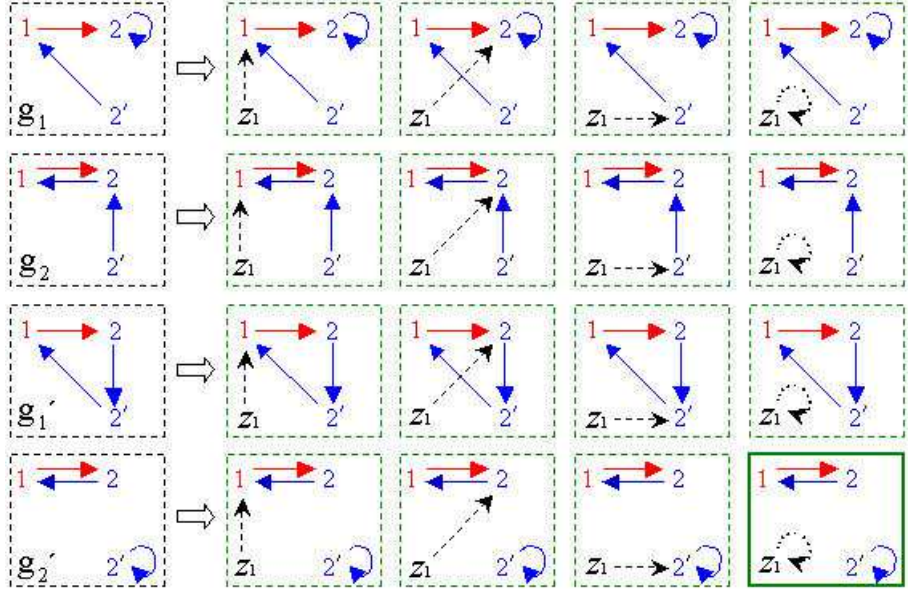}
\end{center}
\caption{The four digraphs in~$\mathcal{G}$ which are constructed from the output digraph in Fig.~\ref{FigDBN1_p5} are shown in the far-left column.  The corresponding pseudo-transition diagrams for each digraph are shown to the right. In Step 4, we have chosen the pseudo-transition diagram in the far-right column which is indicated by the solid border.}
\label{FigDBN2_p5}\end{figure}

In our example, all of the digraphs in~$\mathcal{G}$ are shown in the left-hand column of Fig.~\ref{FigDBN2_p5}. For each digraph, all four of the pseudo-transition diagrams formed from it (by adding one new vertex, $z_{1}$, and assigning just one out-going edge to $z_{1}$) are shown in the columns to the right.

{\bf [Step 4]}  Choose a pseudo-transition diagram $g^{\prime}$ from Step~3, and, as indicated in what follows, use the function $\Xi_{k}$ to label the vertices of~$g^{\prime}$ with the states of a VBN that has $\mu$ nodes. Since $\Xi_{k}$ is not necessarily a one-to-one function, it does not necessarily have an inverse, so randomness will be incorporated into the labeling process.

For every $ l\in\Lambda$, let $S_{kl}=\{s\in S:\Xi_{k}(s)=l\}$.

\begin{itemize}
\item[(i)]      Let $ l\in\Lambda$. If $S_{kl}$ is nonempty and some vertex~$v$ of the output digraph~$g$ (the digraph~$g$ from Step~1, if $k=1$; the digraph~$g$ from the most recent execution of Step~7, if $k>1$) is labeled with~$l$, let $d$ be the out-degree of~$v$ in~$g$.

We show that $d\le|S_{kl}|$: This is obvious if $d=1$, since $S_{kl}$ is nonempty. If $d\ge 2$, then at some point in the construction of our DBN, $v$~was chosen as~$x_{1}$ in Step~2(i), so $v$ has out-going edges to distinct vertices $y_{1},\ldots,y_{d}$ of~$g$. Those vertices have distinct labels in~$g$, so they correspond to distinct states $t_{1},t_{2},\ldots,t_{d}$ in~$S$. Thus there are states $s_{1},s_{2},\ldots,s_{d}$ in~$S_{kl}$ such that the out-going edges from~$v$ to $y_{1},\ldots,y_{d}$ in~$g$ correspond to distinct transitions \[s_{1}\rightarrow t_{1},s_{2}\rightarrow t_{2},\ldots,s_{d}\rightarrow t_{d}\]
 of the VBN whose transition matrix is~$T_{k}$. Then the states $s_{1},\ldots,s_{d}$ are also distinct. (To see this, let $i,j$ be distinct elements of $\{1,2,\ldots,d\}$. Then $t_{i}\ne t_{j}$, and the transitions $s_{i}\rightarrow t_{i}$ and $s_{j}\rightarrow t_{j}$ are distinct, so $s_{i}\ne s_{j}$, since a VBN can make only one transition from a given state.)

If $d=1$, randomly assign some state~$s$ in~$S_{kl}$ as the label of $v$, and let $D_{kl}=\{s\}$. If $d\ge 2$, let $x_{2},\ldots,x_{d}$ be the $d-1$ new vertices added in Step~2(i) when $v$~was chosen as~$x_{1}$, and randomly assign states $s_{1},s_{2},\ldots,s_{d}$ from~$S_{kl}$ as the labels of the vertices $x_{1},\ldots,x_{d}$; then let $D_{kl}=\{s_{1},s_{2},\ldots,s_{d}\}$.\
\item[(ii)]      Randomly assign the states in $S-\bigcup_{l\in\Lambda}\, D_{kl}$ as the labels of the remaining vertices.
\end{itemize}

Let $T$ be the transition matrix of the VBN whose transition diagram was constructed by relabeling the vertices of the pseudo-transition diagram~$g^{\prime}$.

\begin{figure}[h]
\begin{center}
\includegraphics[width=70mm,height=19.6mm]{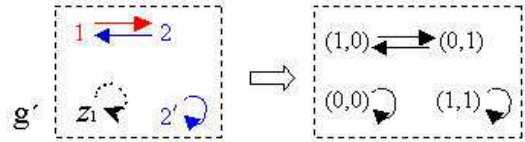}
\end{center}
\caption{Transformation of the chosen pseudo-transition diagram~$g^\prime$ into the transition diagram (of an actual VBN) which is obtained by applying the inverse of the randomly chosen function~$\Xi_1^\prime$ to label the vertices of~$g^\prime$ with the states (of the VBN)}
\label{FigDBN3_p5}\end{figure}

In our example, the following is what transpires in Step~4:

\begin{itemize}
\item[(i)]We assign states to the vertices of~$g^{\prime}$ that are labeled with $1,\ 2$, and $2^{\prime}$. 
\begin{itemize}
\item[$\bullet$]      For $l=1,\ \ S_{1l}=\{(0,0),(1,0)\}$, and the vertex of~$g$ which is labeled with $l$ has out-degree $d=1$.  We randomly assign (1,0) as the label of vertex 1 in~$g^{\prime}$, so $D_{1l}=\{(1,0)\}$.
\item[$\bullet$]   For $l=2,\ S_{1l}=\{(0,1),(1,1)\}$, and the vertex of~$g$ which is labeled with $l$ has out-degree 2. We randomly assign the states $(0,1)$ and $(1,1)$ as the labels of vertices 2 and $2^{\prime}$, respectively, and we have $D_{1l}=\{(0,1),(1,1)\}$.
\item[$\bullet$]   For $l=3$ and $l=4$, we find that $ S_{1l}=\emptyset$, since the image of the labeling function~$\Xi_{1}$ is $\{1,2\}$.
\end{itemize}
\item[(ii)]      We assign a state to the vertex of~$g^{\prime}$ that is labeled with~$z_{1}$. 
\begin{eqnarray*}
S-\bigcup_{l\in\Lambda}D_{1l}&=&S-\left[D_{11}\cup D_{12}\right]\\[.05in]&=&\{(0,0),(0,1),(1,0),(1,1)\}-\left[\{(1,0)\}\cup\{(0,1),(1,1)\}\right]\\[.05in]&=&\{(0,0)\}
\end{eqnarray*}
Hence we assign the state (0,0) as the label of vertex $z_{1}$ in~$g^{\prime}$.

Thus we obtain the following one-to-one function $\Xi_{1}^{\prime}:S\rightarrow\{1,2,2^{\prime},z_{1}\}$:
\begin{equation*}
\Xi_{1}^{\prime}=\left[\begin{array}{cccc}(0,0)&(0,1)&(1,0)&(1,1)\\\hline z_{1}&2&1&2^{\prime}\end{array}\right]
\end{equation*}
\end{itemize}
Applying the inverse of $\Xi_{1}^{\prime}$ to the selected pseudo-transition diagram $g^{\prime}$ from Step~3, we transform~$g^{\prime}$ into the transition diagram of a VBN with 2 nodes, as shown in Fig.~\ref{FigDBN3_p5}.  That is only one of the four possible transition diagrams that we could have obtained---there are two ways to select one state in $S_{1}$ as the label of vertex 1, and two ways to assign the two states in $S_{2}$ as the labels of vertices 2 and $2^{\prime}$.

The transition matrix of the VBN whose transition diagram is shown in Fig.~\ref{FigDBN3_p5}     is
\[T=\left(\begin{array}{cccc}1&0&0&0\\0&0&1&0\\0&1&0&0\\0&0&0&1\\\end{array}\right)\]

{\bf [Step 5]}  Begin the next time step (i.e., increment~$k$ by one unit). Then use the correspondence between the terms of $\vec{S}_{k-2}$ and the terms of~$\vec{\alpha}_{k-1}$ to construct a frequency table as follows (if $k=2$, then $\vec{S}_{k-2}$ ($=S_{0}$) does not exist, so first randomly choose a sequence of states $\vec{S}_{0}$ of a VBN that has $\mu$ nodes): For every pair $(s,l)\in S\times\Lambda$, let $\#(s,l)$ be the frequency of $l$ as the label of~$s$; that is, $\#(s,l)$ is the number of terms~$t$ of $\vec{S}_{k-2}$ such that (a) $s=t$ and (b) $l$ is the corresponding term of~$\vec{\alpha}_{k-1}$. Enter the value of~$\#(s,l)$ in the frequency table.

Once the frequency table is constructed, use it to define the labeling function~$\Xi_{k}:S\rightarrow\Lambda$: Let $s\in S$, and let $m_{s}$ denote the highest frequency associated with~$s$ in~$\vec{\alpha}_{k-1}$, i.e.,\[m_{s}=\max\{\#(s,l):l\in\Lambda\}\]
(Note that $m_{s}=0$ if and only if $s$ is not a term of~$\vec{S}_{k-2}$.) Randomly choose some element of the set $\{l\in\Lambda:\,\#(s,l)=m_{s}\}$ as the value of $\Xi_{k}(s)$. (Note that if $m_{s}=0$, then $\{l\in\Lambda:\,\#(s,l)=m_{s}\}=\Lambda$.)

For time step 2 in our example, we randomly choose $\vec{S}_{0}$  as
\[\vec{S}_{0}=(1,0),(0,0),(0,1),(0,0),(0,1)\]
  Next, we obtain the correspondence between the terms of~$\vec{S}_{0}$ and the terms of~$\vec{\alpha}_{1}$, which is shown in~(\ref{EqCorrespondence1_p5}) and gives rise to the frequency table (Table~\ref{TableFreq_p5}).  

\begin{equation}
\label{EqCorrespondence1_p5}\begin{array}{c|ccccc}\vec{S}_{0}&(1,0)&(0,0)&(0,1)&(0,0)&(0,1)\\\hline\vec{\alpha}_{1}&1&2&1&2&2\end{array}
\end{equation}

\begin{table}[ht]
\begin{center}
\caption{Frequency distribution table derived from the correspondence between $\vec{S}_{0}$ and $\vec{\alpha}_{1}$.}
\begin{tabular}{|c|cccc|}\hline
Label$\backslash$State&(0,0)&(0,1)&(1,0)&(1,1)\\\hline
1&0&1&1&0\\
2&2&1&0&0\\
3&0&0&0&0\\
4&0&0&0&0\\\hline
\end{tabular}
\label{TableFreq_p5}
\end{center}
\end{table}

Then we construct $\Xi_{2}$ from Table~\ref{TableFreq_p5}    as follows: 

\begin{itemize}
\item   $s=(0,0)\Rightarrow m_{s}=2$; there is just one $ l\in\Lambda$ (namely, 2) with $\#(s,l)=m_{s}$, hence $\Xi_{2}(s)=2$.
\item   $s=(0,1)\Rightarrow m_{s}=1$; there are two elements~$l$ of~$\Lambda$ (namely, 1 and 2) with $\#(s,l)=m_{s}$, hence we randomly assign one of them (2) as the value of $\Xi_{2}(s)$.
\item   $s=(1,0)\Rightarrow m_{s}=1$; there is just one $ l\in\Lambda$ (namely, 1) with $\#(s,l)=m_{s}$, hence $\Xi_{2}(s)=1$. 
\item   $s=(1,1)\Rightarrow m_{s}=0$, so $s$ is not a term of~$\vec{S}_{k-2}$, hence we randomly assign an element of~$\Lambda$ (namely, 2) as the value of $\Xi_{2}(s)$.
\end{itemize}   
Thus we obtain 
\begin{equation*}
\Xi_{2}=\left[\begin{array}{cccc}(0,0)&(0,1)&(1,0)&(1,1)\\\hline 2&2&1&2\end{array}\right]
\end{equation*}
This is only one of eight possible labeling functions that we could have used: there are two possibilities for the value of $\Xi_{2}(0,1)$, and four possibilities for the value of $\Xi_{2}(1,1)$.

{\bf [Step 6]}  Let $T_k=T$.

In our example, \[T_2=T=\left(\begin{array}{cccc}1&0&0&0\\0&0&1&0\\0&1&0&0\\0&0&0&1\\\end{array}\right)\]

{\bf [Step 7]}  Determine the sequence of states~$\vec{S}_{k}$ for time step~$k$: Let the first term of~$\vec{S}_{k}$ be the same as the first term of~$\vec{S}_{1}$, and then apply the transition matrix~$T_{k}$ to obtain the remaining terms of~$\vec{S}_{k}$. Once~$\vec{S}_{k}$ has been computed, determine the sequence of labels $\vec{\alpha}_{k}$ which corresponds to~$\vec{S_{k}}$ via the function~$\Xi_{k}$. Then construct the output digraph~$g=(V,E)$ for~$\vec{\alpha}_{k}$ (where $E$ and $V$ are the vertex set and edge set, respectively, of~$g$), and go to Step~2.

\begin{figure}[h]
\begin{center}
\includegraphics[width=23mm,height=8.2mm]{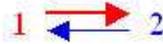}
\end{center}
\caption{Output digraph for the sequence of labels~$\vec{\alpha}_2$}
\label{FigDBN4_p5}
\end{figure}

In our example, the first term of $\vec{S}_{1}$ is $(1,0)$. Applying $T_{2}$, we obtain the sequence of states
\[\vec{S}_{2}=(1,0),(0,1),(1,0),(0,1),(1,0)\]
Then, applying $\Xi_{2}$ to $\vec{S}_{2}$, we obtain the sequence of labels\[\vec{\alpha}_{2}=1,2,1,2,1\]
 and the output digraph for~$\vec{\alpha}_{2}$; the latter is shown in Fig.~\ref{FigDBN4_p5}.

\section{Simulation}\label{SecSimulation_p5}
We performed a number of simulations of a DBN with two nodes. In each simulation, we executed 1,000 trials of 10,000 time steps each. The purpose of each simulation was to investigate the set of rule vectors actually visited by the DBN, which is equivalent to the set of transition matrices actually visited by the DBN.

The number of possible rule vectors for a VBN is\[\left(2^{|S|}\right)^{|V|}=\left(2^{2^{|V|}}\right)^{|V|},\]
 where $S$ is the set of states and $V$ is the node set. Thus a VBN with two nodes has\[\left(2^{2^{2}}\right)^{2}=16^{2}=256\]
possible rule vectors, so at any given time there are 256 possibilities for the rule vector of a DBN with two nodes.

\subsection{Type~1 Simulation}\label{SubSecType1_p5}
In this subsection, we present the results of a simulation of a DBN based on the construction in section~\ref{SecDBN_p5}, which is hereinafter referred to as a type~1 simulation.

\begin{figure}[h]
\begin{center}
\includegraphics[width=75mm,height=56.2mm]{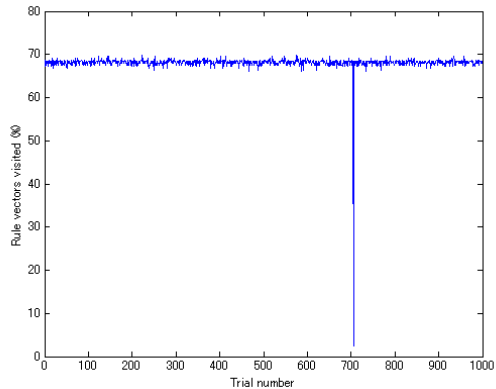}
\end{center}
\caption{Percentage of the 256 rule vectors visited in each of the 1,000 trials of the type~1 simulation}\label{FigResultITI1_p5}\end{figure}

The percentage of the 256 possible rule vectors actually visited in each of the 1,000 trials is shown in Fig.~\ref{FigResultITI1_p5}. Table~\ref{TableResultITI2_p5}         shows the corresponding frequency distribution (the number of trials as a function of the percentage of rule vectors visited).  In every trial, the percentage of rule vectors visited was below 70$\%$. 

\begin{table}[htbp]
\caption{Frequency distribution of the percentage of rule vectors visited in the type~1 simulation}\label{TableResultITI2_p5}
\scriptsize
\begin{tabular}{|c|ccccccc|}            \hline            \nonumber
Percentage of rule&[0,65)& [65,66) & [66,67) & [67,68) & [68,69) & [69, 70) & [70, 100] \\
vectors visited, by interval&&&&&&&\\ \hline
$\#$ of trials among 1,000 & 1 & 0 & 20 & 429 & 493 & 57 & 0 \\ \hline
\end{tabular}
\end{table}

\begin{figure}[h]
\begin{center}
\includegraphics[width=75mm,height=65.6mm]{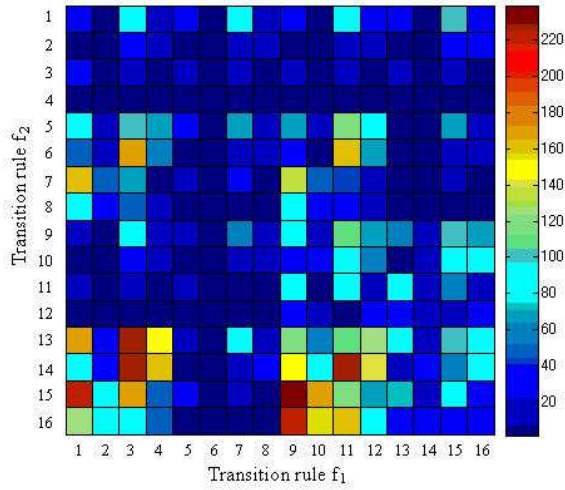}
\end{center}
\caption{Number of time steps at which rule vector $(f_1,f_2)$ was visited in a typical trial of the type~1 simulation with about 68$\%$ coverage}\label{FigResultITI2_p5}\end{figure}

The number of time steps at which rule vector $(f_{1},f_{2})$ was visited in a typical trial of the type~1 simulation with about 68$\%$ coverage (i.e., a trial in which 68$\%$ of the 256 rule vectors were visited) is shown in the map in Fig.~\ref{FigResultITI2_p5}, where $f_{1}$ and $f_{2}$ are the single-node transition rule numbers (from Table~\ref{TableRule_p5}) for nodes~1 and~2, respectively.

\begin{figure}[h]
\begin{center}
\includegraphics[width=75mm,height=58mm]{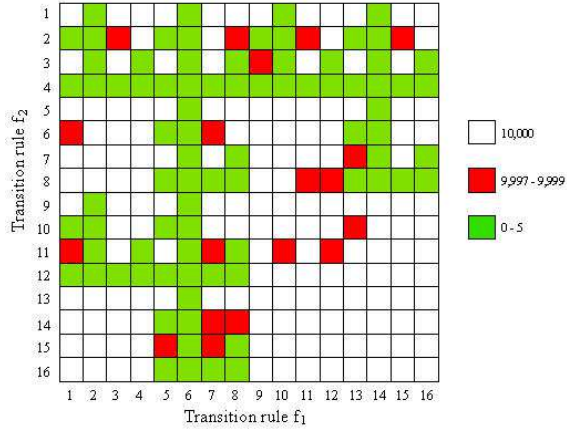}
\end{center}
\caption{Maximum time step at which rule vector~$(f_1,f_2)$ was visited in the first type~1 simulation (where the maximum is taken over all 1,000 trials)}
\label{FigMaxTimeStep1_p5}\end{figure}

The maximum time step at which rule vector $(f_{1},f_{2})$ was visited in the type~1 simulation (where the maximum is taken over all 1,000 trials) is shown in the map in Fig.~\ref{FigMaxTimeStep1_p5}. Of the 256 rule vectors, 156 were visited at the very last (10,000th) time step of at least one trial; those rule vectors are depicted in white in the figure. Red is used for each of the 19 rule vectors that were visited at some time step in the range 9,997--9,999 but were never visited at the last time step. The remaining 81 rule vectors, which are shown in green, were not visited after the fifth time step of any trial. Each of those 81 rule vectors was visited in at most 21 of the 1,000 trials, and none of them was visited in the 943 trials with less than 69$\%$ coverage. Thus we ran two additional type 1 simulations: one in which we randomly chose the initial rule vector in each trial from the other 175 ($=256-81$) rule vectors, and one in which we eliminated the data for the first five time steps.  In those two simulations, none of the aforementioned 81 rule vectors was visited at all---in contrast to the other 175 rule vectors, each of which was visited in at least 877 of the 1,000 trials.

\begin{figure}[h]
\begin{center}
\includegraphics[width=130mm,height=57.8mm]{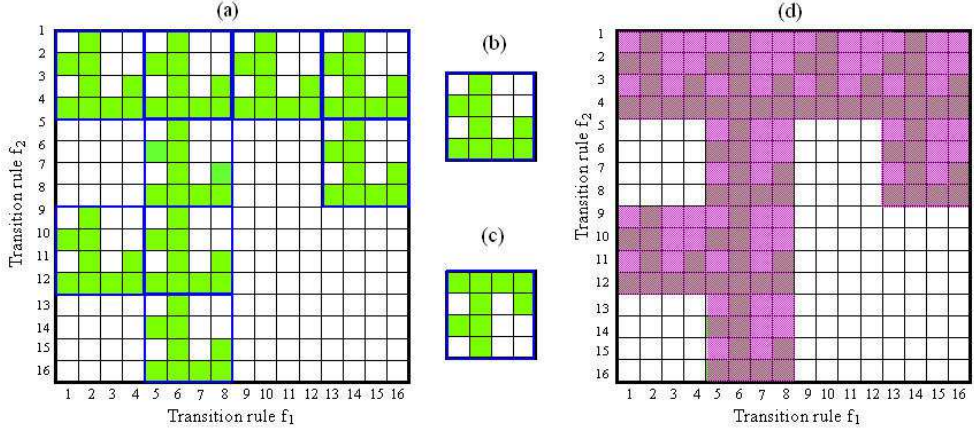}
\end{center}
\caption{(a) The 81 rule vectors (indicated by green squares) not visited after the fifth time step in any trial of the first type 1 simulation. (b) Configuration of the nine green squares in each of the nine partially green $4\times 4$ blocks in (a). (c) Block in (b) turned upside down. (d) Configuration of the nine partially green blocks in (a), which is identical to the configuration of the nine green squares in (c).}\label{FigResultITI3_p5}\end{figure}

The 81 rule vectors not visited after the fifth time step in the first type~1 simulation are depicted by green squares in Fig.~\ref{FigResultITI3_p5}(a). They include rule vector $(4,f_{2})$ for every~ transition rule~$f_{2}$, and rule vector $(f_{1},6)$ for every~ transition rule $f_{1}$.  

From Table~\ref{TableRule_p5}, there are four single-node transition rules with exactly one virtual incoming node (namely, rules 4, 6, 11, and 13).  Note the following properties of those four rules:
\begin{enumerate}
\item   Rule 4 is the negation of rule 13, and rule 6 is the negation of rule 11.
\item   Rules 11 and 13 are rules of an LVBN, but rules 4 and 6 are not.
\end{enumerate}

Fig.~\ref{FigResultITI3_p5}(a) is partitioned into sixteen $4\times 4$ blocks, nine of which are colored identically to the one shown in Fig.~\ref{FigResultITI3_p5}(b); the remaining seven blocks are totally white.  Furthermore, if the block in Fig.~\ref{FigResultITI3_p5}(b) is turned upside down, as shown in Fig.~\ref{FigResultITI3_p5}(c), the configuration of the nine green squares within it is identical to the configuration of the nine partially green blocks within Fig.~\ref{FigResultITI3_p5}(a). That relationship can be seen more readily by comparing Figs.~\ref{FigResultITI3_p5}(c) and~\ref{FigResultITI3_p5}(d).

\begin{figure}[h]
\begin{center}
\includegraphics[width=75mm,height=60mm]{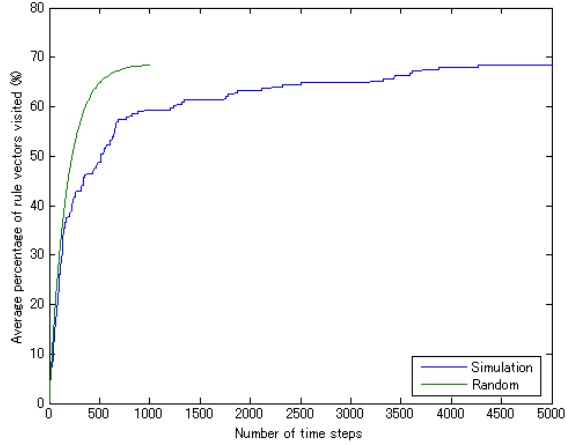}
\end{center}
\caption{Blue curve: graph of the cumulative percentage of rule vectors visited in the first type~1 simulation, averaged over the 550 trials with at least 68\% coverage, vs.\ the number of time steps.  Shown in green is the parametric curve $\displaystyle\left(\theta(175,m),\,\frac{m}{256}\cdot 100\right)$, where $m$ varies from 1 to 175 and $\theta(175,m)$ is given by (\ref{EqVisitratingfunction1_p5}).}\label{FigResultITI4_p5}
\end{figure}

Fig.~\ref{FigResultITI4_p5}     shows a graph of the cumulative percentage of rule vectors visited in our first type 1 simulation, averaged over the 550 trials in which at least 68$\%$ of the 256 rule vectors were visited (i.e., the trials in which at least 175 rule vectors were visited), vs.\ the number of time steps.  The blue curve is the result of our simulation. Shown in green is the parametric curve $\displaystyle \left(\theta(175,m),\,\frac{m}{256}\cdot 100\right)$, where $\theta(175,m)$ is the (theoretical) expectation value of the number of time steps taken to visit $m$ rule vectors out of a given set of 175 possible rule vectors (the number of rule vectors that were last visited at time step~9,997---or later---in some trial of our simulation) and $m$ varies from 1 to 175:

\begin{equation}
\label{EqVisitratingfunction1_p5}
\displaystyle \theta(175,m)=\sum_{i=1}^{m}\frac{175}{175-i+1}
\end{equation}
The quantity $\displaystyle \frac{m}{256}\cdot 100$ is the value of $m$ expressed as a percentage of 256 (the total number of possible rule vectors for a DBN with two nodes). The formula in (\ref{EqVisitratingfunction1_p5}) is obtained by assuming that, at every time step, each of the 175 possible rule vectors has probability $\displaystyle \frac{1}{175}$ of being visited. The expected number of time steps taken to visit 175 rule vectors is $\theta(175,175)\approx$1,005.3.  However, the number of time steps actually taken to visit an average of 175 rule vectors (in the 550 trials of our type~1 simulation with at least 68$\%$ coverage) was about 5,000.

\subsection{Type~2 Simulation}\label{SubSecType2_p5}
In our remaining simulations, we used a DBN construction which is similar to the one presented in Section~\ref{SecDBN_p5}, the only difference being that, for $k\ge 2$, we chose the transition matrix $T_{k}$ of the DBN for time step~$k$ to be $(Q_{k})^{-1}TQ_{k}$, where $T$ is the transition matrix from Step~4 of our original construction and $Q_{k}$ is the matrix representation of some permutation $P_{k}$ of the $2^{\mu}$ states of a DBN with $\mu$ nodes (in the case of our simulations, $P_{k}$~was some permutation of the four states of a DBN with two nodes). We ran three different types of such simulations, corresponding to three different ways of choosing the permutations.  

In this subsection, we present the results of our type~2 simulation, in which $P_{k}$ (for $k\ge 2$) was randomly chosen from the following set of permutations: 
\[\{e\}\ \bigcup\ \{(1\ 2),(1\ 3),(1\ 4),(2\ 3),(2\ 4),(3\ 4)\}\]
\[\hspace*{1in}\bigcup\ \{(1\ 2)(3\ 4),(1\ 3)(2\ 4),(1\ 4)(2\ 3)\}\]
Here, $e$ denotes the identity permutation. Every permutation of the form $(a\ b)$ denotes a transposition (a single 2-cycle), and every permutation of the form $(a\ b)(c\ d)$ denotes the product of a disjoint pair of 2-cycles.

\begin{figure}[h]
\begin{center}
\includegraphics[width=75mm,height=59.4mm]{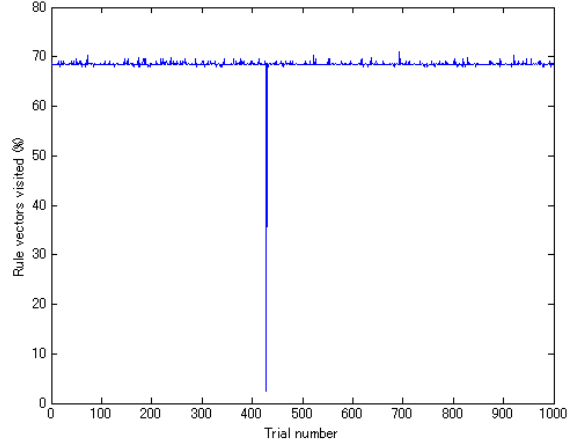}
\end{center}
\caption{Percentage of the 256 rule vectors visited in each of the 1,000 trials of the type~2 simulation}\label{FigResultType2_1_p5}
\end{figure}

The percentage of the 256 possible rule vectors actually visited in each of the 1,000 trials of our type~2 simulation is shown in Fig.~\ref{FigResultType2_1_p5}. Table~\ref{TableResultType2_2_p5}        shows the corresponding frequency distribution.  In every trial, the percentage of rule vectors visited was below 71$\%$. 

\begin{table}[htbp]
\caption{Frequency distribution of the percentage of rule vectors visited in the type~2 simulation}\label{TableResultType2_2_p5}
\scriptsize
\begin{tabular}{|c|ccccc|}          \hline          \nonumber
Percentage of rule &[0,67) & [67,68) & [68,69) & [69, 70) & [70, 71) \\
vectors visited, by interval&&&&&\\ \hline
$\#$ of trials among 1,000 & 1 & 33 & 877 & 85 & 4 \\ \hline
\end{tabular}
\end{table}

\begin{figure}[h]
\begin{center}
\includegraphics[width=75mm,height=65.8mm]{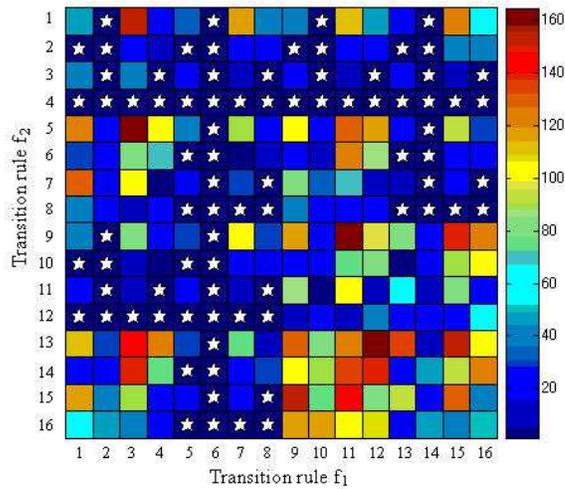}
\end{center}
\caption{Number of time steps at which rule vector $(f_1,f_2)$ was visited in a typical trial of the type~2 simulation with about 68$\%$ coverage}\label{FigResultType2_3_p5}
\end{figure}

The number of time steps at which rule vector $(f_{1},f_{2})$ was visited in a typical trial of the type~2 simulation with about 68$\%$ coverage is shown in the map in Fig.~\ref{FigResultType2_3_p5}.  There and in each map presented later in this paper, the 81 rule vectors from Fig.~\ref{FigResultITI3_p5}(a) are marked with a star. The associated map of the maximum time step at which each rule vector was visited in the type~2 simulation is very similar to the map shown in Fig.~\ref{FigMaxTimeStep1_p5}  for the type~1 simulation, so we have not presented it here.

None of the 81 rule vectors from Fig.~\ref{FigResultITI3_p5}(a) was visited after the sixth time step in any trial of our type~2 simulation, and each of those 81 rule vectors was visited in at most 14 of the 1,000 trials. Thus we ran two additional type~2 simulations: one in which we randomly chose the initial rule vector in each trial from the other 175 rule vectors, and one in which we eliminated the data for the first six time steps.  As in type~1, none of those 81 rule vectors was visited at all in the two additional simulations---in contrast to the other 175 rule vectors, each of which was visited in at least 880 of the 1,000 trials.

\begin{figure}[h]
\begin{center}
\includegraphics[width=75mm,height=59.4mm]{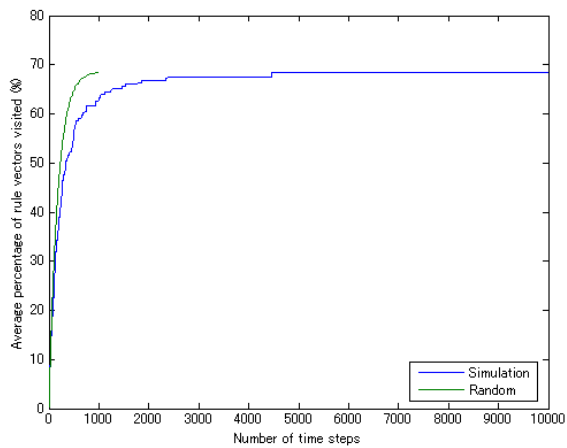}
\end{center}
\caption{Blue curve: graph of the cumulative percentage of rule vectors visited in the first type~2 simulation, averaged over the 966 trials with at least 68\% coverage, vs.\ the number of time steps.  Shown in green is the parametric curve $\displaystyle\left(\theta(175,m),\,\frac{m}{256}\cdot 100\right)$, where $m$ varies from 1 to 175 and $\theta(175,m)$ is given by (\ref{EqVisitratingfunction1_p5}).}\label{FigResultType2_4_p5}
\end{figure}

Fig.~\ref{FigResultType2_4_p5}     shows a graph of the cumulative percentage of rule vectors visited in our first type 2 simulation, averaged over the 966 trials in which at least 68$\%$ of the 256 rule vectors were visited (i.e., the trials in which at least 175 rule vectors were visited), vs.\ the number of time steps.  Again, the blue curve is the result of our simulation, and the green curve is the parametric curve $\displaystyle \left(\theta(175,m),\,\frac{m}{256}\cdot 100\right)$. The number of time steps actually taken to visit an average of 175 rule vectors (in the 966 trials of our type~2 simulation with at least 68$\%$ coverage) was about 4,400, which is somewhat smaller than the 5,000 or so steps taken in our type~1 simulation---and closer to the (theoretical) expectation value of about~1,005.3. This difference in the results of the two simulations could possibly be due to the use of randomness in selecting the permutation~$P_{k}$ that figures in the computation of the transition matrix~$T_{k}$ (for time step~$k$) in the type~2 simulation.

\subsection{Type~3 Simulation}\label{SubSecType3_p5}
In this subsection, we present the results of our type~3 simulation, in which $P_{k}$ (for $k\ge 2$) was randomly chosen from the set of all 24 permutations of the four states of a DBN with two nodes. 

Each of the 256 rule vectors was visited in all 1,000 trials of our type~3 simulation, and the 81 rule vectors from Fig.~\ref{FigResultITI3_p5}(a) were visited much more frequently in the type~3 simulation than in the type~1 and type~2 simulations. In fact, the total number of visits per rule vector (summed over all 1,000 trials of the type~3 simulation) was just under two-thirds as large, on average, for those 81 rule vectors as for the other 175.

\begin{figure}[h]
\begin{center}
\includegraphics[width=75mm,height=66.6mm]{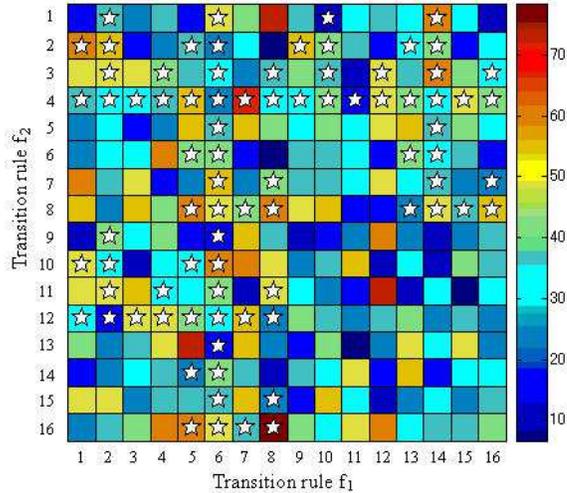}
\end{center}
\caption{Number of time steps at which rule vector $(f_1,f_2)$ was visited in a typical trial of the type~3 simulation (with 100$\%$ coverage)}\label{FigResultType3_3_p5}
\end{figure}

The number of time steps at which rule vector $(f_{1},f_{2})$ was visited in a typical trial of the type~3 simulation is shown in the map in Fig.~\ref{FigResultType3_3_p5}.

\begin{figure}[h]
\begin{center}
\includegraphics[width=80mm,height=62.8mm]{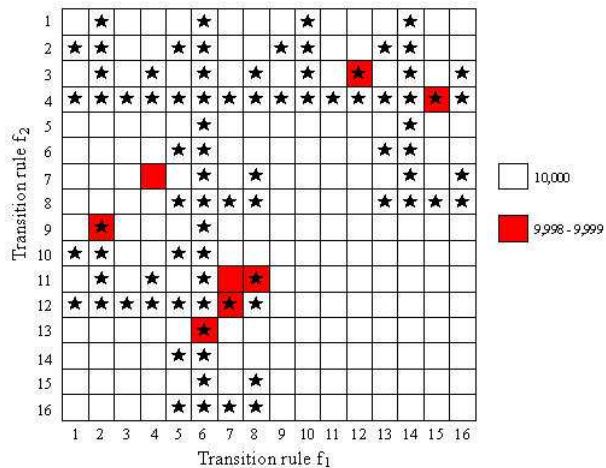}
\end{center}
\caption{Maximum time step at which rule vector~$(f_1,f_2)$ was visited in the type~3 simulation (where the maximum is taken over all 1,000 trials)}
\label{FigMaxTimeStep3_p5}\end{figure}

The maximum time step at which rule vector $(f_{1},f_{2})$ was visited in the type~3 simulation (where the maximum is taken over all 1,000 trials) is shown in the map in Fig.~\ref{FigMaxTimeStep3_p5}. All but 8 of the 256 rule vectors were visited at the very last (10,000th) time step of at least one trial$;$ those 248 rule vectors are depicted in white in the figure. Red is used for the other 8 rule vectors, each of which was visited at the 9,998th or 9,999th time step of at least one trial.

\begin{figure}[h]
\begin{center}
\includegraphics[width=75mm,height=58.6mm]{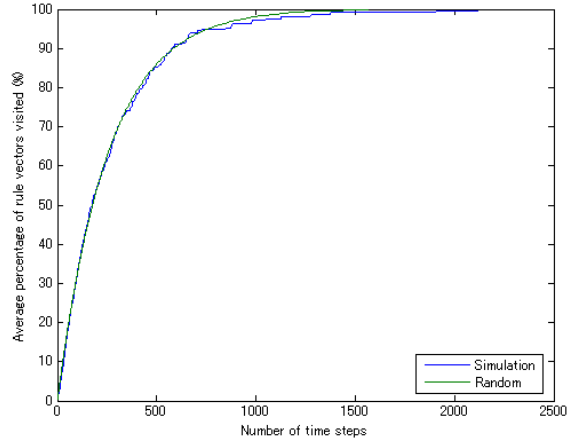}
\end{center}
\caption{Blue curve: graph of the cumulative percentage of rule vectors visited in the type~3 simulation, averaged over all 1,000 trials, vs.\ the number of time steps.  Shown in green is the parametric curve $\displaystyle\left(\theta(256,m),\,\frac{m}{256}\cdot 100\right)$, where $m$ varies from 1 to 256 and $\theta(256,m)$ is given by (\ref{EqVisitratingfunction2_p5}).}\label{FigResultType3_4_p5}
\end{figure}

Fig.~\ref{FigResultType3_4_p5}  shows a graph of the cumulative percentage of rule vectors visited in our type 3 simulation, averaged over all 1,000 trials, vs.\ the number of time steps.  The blue curve is the result of our simulation, and the green curve is the parametric curve $\displaystyle \left(\theta(256,m),\,\frac{m}{256}\cdot 100\right)$, where $\theta(256,m)$ is the (theoretical) expectation value of the number of time steps taken to visit $m$ rule vectors (out of the set of all 256 possible rule vectors) and $m$ varies from 1 to 256:
\begin{equation}
\label{EqVisitratingfunction2_p5}
\displaystyle \theta(256,m)=\sum_{i=1}^{m}\frac{256}{256-i+1}
\end{equation}
The formula in (\ref{EqVisitratingfunction2_p5}) is obtained by assuming that, at every time step, each of the 256 possible rule vectors has probability $\displaystyle \frac{1}{256}$ of being visited. The expected number of time steps taken to visit 256 rule vectors is $\theta(256,256)\approx$1567.8, and the number of time steps actually taken 
to visit all 256 rule vectors (in all 1,000 trials of our type~3 simulation) was about 1,650.

We also ran a simulation in which $P_{k}$ was randomly chosen from  the complement of the set of permutations used in type~2:
\[\{(1\ 2\ 3),(1\ 2\ 4),(1\ 3\ 2),(1\ 3\ 4),(1\ 4\ 2),(1\ 4\ 3),(2\ 3\ 4),(2\ 4\ 3)\}\]
\[\hspace*{1in}\bigcup\{(1\ 2\ 3\ 4),(1\ 2\ 4\ 3),(1\ 3\ 2\ 4),(1\ 3\ 4\ 2),(1\ 4\ 2\ 3),(1\ 4\ 3\ 2)\}\]
That is, $P_{k}$ was chosen to be either a 3-cycle or a 4-cycle. The results were similar to those for our type~3 simulation, so we have not presented them here.

\subsection{Type~4 Simulation}\label{SubSecType4_p5}
In this subsection, we present the results of our type~4 simulation, in which permutation $P_{k}$ (for $k\ge 2$) was constructed from the labeling function~\mbox{$\Xi_{k}:S\rightarrow\Lambda$}, as follows:

\begin{itemize}
\item[(i)]         For every $ l\in\Lambda$, let $S_{kl}=\{s\in S:\Xi_{k}(s)=l\}$ and $r_{kl}=|S_{kl}|$.
\item[(ii)]         For every $l$ with $r_{kl}>1$, randomly choose a permutation $p_{kl}$ of the states in~$S_{kl}$ such that $p_{kl}$ is an $r_{kl}$-cycle. Then let $P_{k}$ consist of the product of all the cycles thus chosen:\[P_{k}=\prod_{\begin{array}{c}\\[-.4in]l\in\Lambda\\[-.15in]r_{kl}>1\end{array}}\hspace{-.1in}p_{kl}\]
\end{itemize}
In our example, the following is what transpires at time step~2 in constructing~$P_{2}$ from the labeling function
\begin{equation*}
\Xi_{2}=\left[\begin{array}{cccc}(0,0)&(0,1)&(1,0)&(1,1)\\\hline 2&2&1&2\end{array}\right]
\end{equation*}
\begin{itemize}
\item[(i)]$\\$
\begin{itemize}
\item[$\bullet$]           For $l=1,\ \ S_{2l}=\{(1,0)\}$. Thus $r_{2l}=1$, 
so no permutation is chosen for $l=1$.
\item[$\bullet$]For $l=2,\ \ S_{2l}=\{(0,0),(0,1),(1,1)\}$. Thus $r_{2l}=3$, so we randomly choose a 3-cycle $p_{2l}$ of the three states in~$S_{2l}$. In the lexicographical ordering scheme, they are states 1, 2, and 4, so there are two such 3-cycles to choose from: $(1$\,$2$\,$4)$ and $(1$\,$4$\,$2)$. We choose $(1$\,$4$\,$2)$ as $p_{2l}$.
\item[$\bullet$]For $l=3$ and $ l=4,\ \ S_{2l}=\emptyset$, so no permutation is chosen for either of them.
\end{itemize}
\item[(ii)]           We obtain the permutation $P_{2}=(1$\,$4$\,$2)$. The matrix representation of~$P_{2}$ is 
\[Q_{2}=\left(\begin{array}{cccc}0&0&0&1\\1&0&0&0\\0&0&1&0\\0&1&0&0\\\end{array}\right)\]
\end{itemize}        
Thus we obtain the transition matrix\[T_{2}=(Q_{2})^{-1}TQ_{2}=\left(\begin{array}{cccc}0&1&0&0\\0&0&0&1\\0&0&1&0\\1&0&0&0\\\end{array}\right)\cdot\left(\begin{array}{cccc}1&0&0&0\\0&0&1&0\\0&1&0&0\\0&0&0&1\\\end{array}\right)\cdot\left(\begin{array}{cccc}0&0&0&1\\1&0&0&0\\0&0&1&0\\0&1&0&0\\\end{array}\right)=\left(\begin{array}{cccc}0&0&1&0\\0&1&0&0\\1&0&0&0\\0&0&0&1\\\end{array}\right)\]

\begin{figure}[h]
\begin{center}
\includegraphics[width=75mm,height=59.4mm]{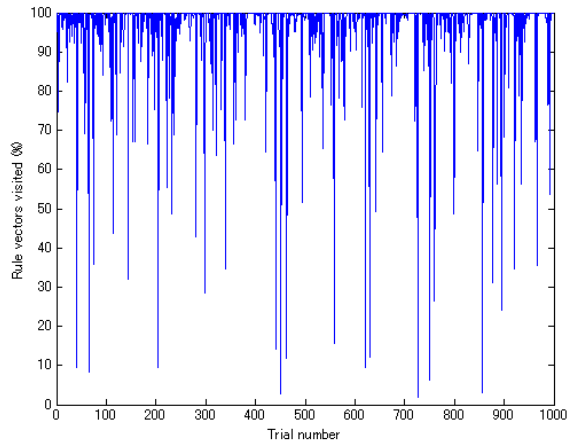}
\end{center}
\caption{Percentage of the 256 rule vectors visited in each of the 1,000 trials of the type~4 simulation}\label{FigResultType4_1_p5}
\end{figure}

The percentage of the 256 possible rule vectors actually visited in each of the 1,000 trials of our type~4 simulation is shown in Fig.~\ref{FigResultType4_1_p5}. Table~\ref{TableResultType4_2_p5}  shows the corresponding frequency distribution.  In 906 of the 1,000 trials, at least 90$\%$ of the rule vectors were visited; also, every rule vector was visited in at least 775 trials. Furthermore, the total number of visits per rule vector (summed over all 1,000 trials of the type~4 simulation) was nearly twice as large, on average, for the 81 rule vectors from Fig.~\ref{FigResultITI3_p5}(a) than for the other 175 rule vectors. 

\begin{table}[htbp]
\caption{Frequency distribution of the percentage of rule vectors visited in the type~4 simulation}\label{TableResultType4_2_p5}
\scriptsize
\begin{tabular}{|c|ccccccc|}        \hline        \nonumber
Percentage of rule&[0,50)& [50,60) & [60,70) & [70,80) & [80,90) & [90,100) & 100 \\
vectors visited, by interval&&&&&&&\\ \hline
$\#$ of trials among 1,000  & 26 & 5 & 16 & 20 & 27 & 356 & 550 \\ \hline
\end{tabular}
\end{table}

The 1,000 trials of the type~4 simulation can be grouped into two classes:
\begin{itemize}
\item[\textbf{Class (i)}]  trials in which a handful of rule vectors were visited very frequently (visited at roughly 1,000 or more different time steps) and the remaining rule vectors were visited very infrequently (visited at roughly no more than 30 different time steps)
\item[\textbf{Class (ii)}]  trials in which the variation in the number of visits per rule vector was very small
\end{itemize}
About 70$\%$ of the trials of the type~4 simulation fall into class (i); the remaining 30$\%$ are in class (ii). All 94 trials with less than 90$\%$ coverages are in class (i).

\begin{figure}[h]
\begin{center}
\includegraphics[width=130mm,height=60.6mm]{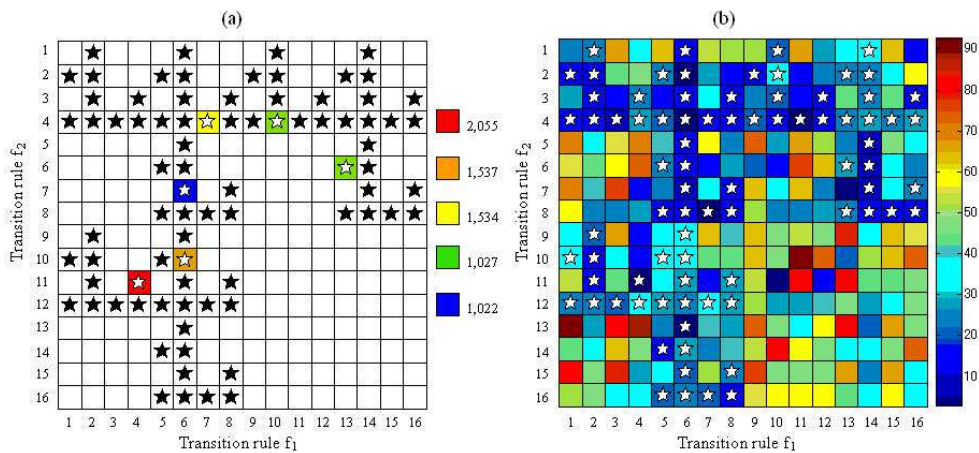}
\end{center}
\caption{Number of time steps at which rule vector $(f_1,f_2)$ was visited in two trials of the type~4 simulation with at least 90\% coverage apiece: (a) a typical class (i) trial and (b) a typical class (ii) trial}
\label{FigResultType4_3_p5}
\end{figure}

The numbers of time steps at which rule vector $(f_{1},f_{2})$ was visited in two specific trials of the type~4 simulation with at least 90$\%$ coverage apiece are shown in the maps in Fig.~\ref{FigResultType4_3_p5}. The map in (a) is for a typical class (i) trial, and the map in (b) is for a typical class (ii) trial.

Only six rule vectors (namely, $(4,11),\,(6,7),\,(6,10),\,(7,4),\,(10,4)$, and $(13,6)$) were visited very frequently (visited at roughly 1,000 or more different time steps) in the class~(i) trial associated with Fig.~\ref{FigResultType4_3_p5}(a). White is used for the remaining 250 rule vectors, each of which was visited no more than 22 times. In about 20$\%$ of the class~(i) trials, at least one rule vector other than those six was among the rule vectors visited very frequently (visited at roughly 1,000 or more different time steps). 

The map shown in Fig.~\ref{FigResultType4_3_p5}(b), for a class (ii) trial, bears a much greater resemblance to the maps for the trials of the type~3 simulation than to those for the class (i) trials of the type~4 simulation.

All six of the rule vectors enumerated earlier are from Fig.~\ref{FigResultITI3_p5}(a). Each of those six rule vectors was visited a total of more than 300,000 times in the class~(i) trials of the type~4 simulation, while each of the 175 rule vectors not from Fig.~\ref{FigResultITI3_p5}(a) had a total of fewer than 50,000 visits in the class~(i) trials.

\begin{figure}[h]
\begin{center}
\includegraphics[width=130mm,height=65mm]{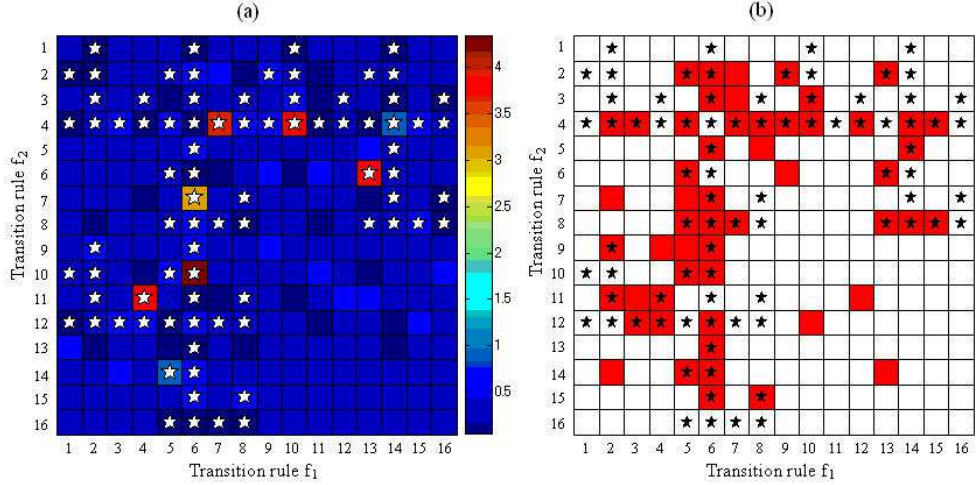}
\end{center}
\caption{(a) Total number of visits (in units of $10^5$) of rule vector $(f_1,f_2)$ in the class (i) trials of the type~4 simulation. (b) The 54 rule vectors (indicated by red squares) that were visited very frequently in at least one of the class (i) trials.}\label{FigBC_p5}\end{figure}

The total number of visits (in units of $10^{5}$) of rule vector $(f_{1},f_{2})$ in the class (i) trials of the type~4 simulation (summed over all 1,000 trials) is shown in the map in Fig.~\ref{FigBC_p5}(a).

Only 54 of the 256 rule vectors were visited very frequently in at least one of the class~(i) trials: 41 of the 81 rule vectors from Fig.~\ref{FigResultITI3_p5}(a), including the six rule vectors enumerated earlier; and 13 of the other 175 rule vectors. Those 54 rule vectors are depicted by red squares in Fig.~\ref{FigBC_p5}(b).

\begin{figure}[h]
\begin{center}
\includegraphics[width=80mm,height=63.2mm]{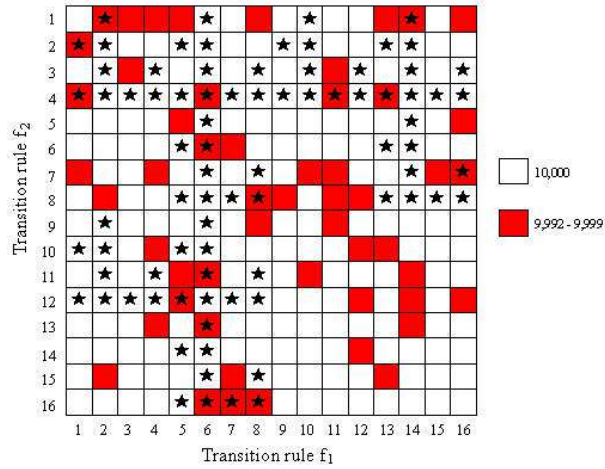}
\end{center}
\caption{Maximum time step at which rule vector~$(f_1,f_2)$ was visited in the type~4 simulation (where the maximum is taken over all 1,000 trials)}
\label{FigMaxTimeStep4_p5}\end{figure}

The maximum time step at which rule vector $(f_{1},f_{2})$ was visited in the type~4 simulation (where the maximum is taken over all 1,000 trials) is shown in the map in Fig.~\ref{FigMaxTimeStep4_p5}. Of the 256 rule vectors, 203 were visited at the very last (10,000th) time step of at least one trial; those rule vectors are depicted in white in the figure. Red is used for the other 53 rule vectors, each of which was visited at some time step in the range 9,992--9,999 but was never visited at the last time step.

\begin{figure}[h]
\begin{center}
\includegraphics[width=75mm,height=59.2mm]{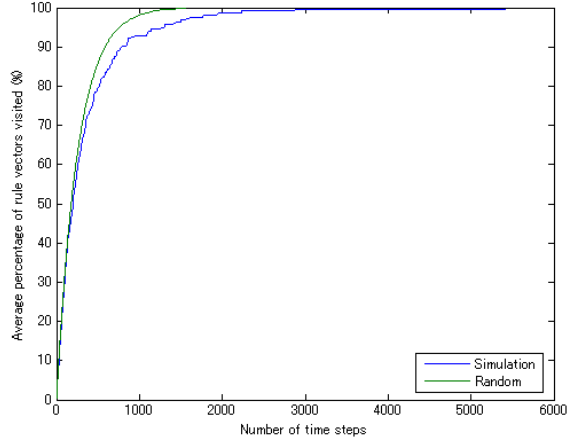}
\end{center}
\caption{Blue curve: graph of the cumulative percentage of rule vectors visited in the type~4 simulation, averaged over the 550 trials with 100\% coverage, vs.\,the number of time steps.  Shown in green is the parametric curve $\displaystyle\left(\theta(256,m),\,\frac{m}{256}\cdot 100\right)$, where $m$ varies from 1 to 256 and $\theta(256,m)$ is given by (\ref{EqVisitratingfunction2_p5}).}\label{FigResultType4_4_p5}
\end{figure}

Fig.~\ref{FigResultType4_4_p5}  shows a graph of the cumulative percentage of rule vectors visited in our type 4 simulation, averaged over the 550 trials in which all 256 rule vectors were visited, vs.\ the number of time steps. Again, the blue curve is the result of our simulation, and the green curve is the parametric curve given by $\displaystyle \left(\theta(256,m),\,\frac{m}{256}\cdot 100\right)$. The number of time steps actually taken to visit all 256 rule vectors (in the 550 trials of our simulation with 100$\%$ coverage) was about 4,500, which is about 2.9 times as large as the theoretical value and 2.7 times the number of steps taken in the type~3 simulation. 

\section{Concluding Remarks}\label{SecRemarks_p5}
At each time step in our simulations of a DBN, we incorporated randomness into several processes: in generating pseudo-transition diagrams from the output digraph, in choosing a pseudo-transition diagram and then transforming it to the transition diagram of a VBN, in choosing a labeling function for the states of the DBN, and (in types~2, 3, and~4) in choosing a permutation of the states of the DBN.

In every trial of the type~1 simulation, fewer than 70$\%$ of the rule vectors were visited. Furthermore, the 256 rule vectors split into two groups: the set of 81 rule vectors that were not visited after the fifth time step in any trial of the first type~1 simulation (and not visited at all in the two additional type~1 simulations), and the set consisting of the other 175 rule vectors, each of which was visited considerably more frequently than those in the set of 81. Those features of the type~1 simulation were largely preserved under the kinds of permutations of the states of the DBN which were chosen at random in the type~2 simulation (namely, the identity permutation, the 2-cycles, and the products of disjoint pairs of 2-cycles).

The salient features of the type~1 and type~2 simulations were not preserved under the kinds of permutations chosen in the type~3 and type~4 simulations.  (Any permutation could be selected in type~3; however, the permutation $P_{k}$ selected at time step~$k$ in the type~4 simulation depended in part on the labeling function $\Xi_{k}$.)  A remarkable contrast was found between the results of the type~3 and type~4 simulations on one hand, and those for type~1 and type~2 on the other. All 256 rule vectors were visited in every trial of the type~3 simulation, and six of the 81 rule vectors which were not visited after the sixth time step of any trial in the type~1 and type~2 simulations were the rule vectors visited the most frequently of all---by far---in the type~4 simulation. 

The results of the one simulation we ran in which $P_{k}$ was randomly chosen from the complement of the set of permutations used in type~2 were similar to those for our type~3 simulation, in that the total number of visits per rule vector varied little from one rule vector to another. 

Fully 41 of the aforementioned 81 rule vectors were visited very frequently (visited at roughly 1,000 or more different time steps) in at least one class~(i) trial of our type~4 simulation. Six of those 41 rule vectors had a total of over 300,000 visits in the type~4 simulation. 

Each component of every one of those six rule vectors is a single-node transition rule from the set $\{4,6,7,10,11,13\}$. Three of those (rule numbers 7, 11, and 13) are transition rules of an LVBN; each of the other three (4, 6, and 10) is the negation of a single-node transition rule of an LVBN. The only single-node transition rules of an LVBN which are not a component of any of those six rule vectors are rule numbers 1 and 16---and both of those transition rules correspond to a node of an LVBN that has no virtual incoming nodes.  Thus it appears that we can regard the class (i) trials of the type~4 simulation as a restriction of the single-node transition rules of a VBN to the single-node transition rules of an LVBN and their negations, hence that linearity may have played a significant role in our type~4 simulation.  Therefore, linearity may be a worthwhile topic for further study.



\begin{thebibliography}{99}
\bibitem{Aczel1988}     Aczel, P., ``Non-well-founded Sets,'' Stanford CSLI 1988.

\bibitem{BarwiseEtchemendy1987}     Barwise, J., and Etchemendy, J. {\it The Liar}: {\it An Essay on Truth and Circularity}, Oxford University Press (New York), 1987.

\bibitem{ChemeroTurvey2006}     Chemero, A., and Turvey, M.T., ``Complexity and Closure to Efficient Cause,'' Proceedings of AlifeX: Workshop on Artificial Autonomy, 2006.

\bibitem{GunjiItoKusunoki}     Gunji, P.-Y., Ito, K., and Kusunoki, Y.,  ``Formal Model of Internal Measurement: Alternate Changing between Recursive Definition and Domain Equation,''  Physica D 110 (1997), 289--312.

\bibitem{GunjiHarunaSawa}     Gunji, P.-Y., Haruna, T., and Sawa, K., ``Principles of biological organization: Local-global negotiation based on material cause,'' Physica D 219 (2006), Issue 2, 152--167.


\bibitem{Kercel2003}     Kercel, S.W., ``Endogenous causes-bizarre effects,'' Evolution and Cognition 8 (2003), 130--144.

\bibitem{Kauffman1}     Kauffman, S.A., ``Homeostasis and Differentiation in random Genetic Control Networks,''  Nature, 224 (1969) 177--178.

\bibitem{Kauffman2}     Kauffman, S.A., and Grass, K., ``The Logical Analysis of Continuous, Nonlinear Biochemical Control Networks,''  J. Theoretical Biology, 39 (1973), 103--129.

\bibitem{Kauffman3}     Kauffman, S.A., ``The Large Scale Structure and Dynamics of Genetic Control Circuits: An Ensemble Approach,''  J. Theoretical Biology, 44 (1974), 167--190.

\bibitem{Lawvere1969}     Lawvere, F. W., ``Diagonal Arguments and Cartesian Closed Categories,'' In {\it Lecture Notes in Mathematics No. 92} (1969), Springer Verlag, Berlin, 134--145.

\bibitem{MacLane}     MacLane, S., {\it Categories for Working Mathematicians}, Springer, Berlin (1971).

\bibitem{Matsuno1}     Matsuno, K., {\it Protobiology}: {\it Physical Basis of Biology}, CRC Press, Boca Raton, FL, 1989.

\bibitem{Matsuno2}     Matsuno, K., ``Forming and maintaining a heat engine for quantum biology,'' BioSystems 85 (2006) 23--29.

\bibitem{Rosen1}     Rosen, R., ``Some Relational Cell Models: The Metabolism--Repair Systems,'' In {\it Foundation of Mathematical Biology}, Vol. II (1972) 217--253, Academic Press, New York.

\bibitem{Rosen2}     Rosen, R., ``Theoretical Biology and Complexity,'' In {\it Three Essays on the Natural Philosophy of Complex Systems},  Academic Press, London, 1985.

\bibitem{Rosen3}     Rosen, R., {\it Essays on Life Itself}, Columbia University Press, New York, 2000.

\bibitem{Shmulevich1}     Shmulevich, I., Dougherty, E.R., Kim, S., and Zhang, W. ``Probabilistic Boolean Networks: A Rule-based Uncertainty Model for Gene Regulatory Networks. Bioinformatics,'' 18(2) (2002), 261--274. 

\bibitem{Shmulevich2}     Shmulevich, I., Dougherty, E.R., and Zhang, W. ``From Boolean to Probabilistic Boolean Networks as Models of Genetic Regulatory Networks. Proceedings of the IEEE,'' 90(11) (2002), 1778--1792. 

\end{thebibliography}
\end{document}